\pdfoutput=1
\documentclass[12pt]{article}
\usepackage[
top = 2.5cm, 
bottom = 2.5cm,  
left = 2.55cm,  
right = 2.55cm]{geometry}

\usepackage{subfigure}
\usepackage{caption}
\usepackage{latexsym} 
\usepackage{verbatim} 
\usepackage{tikz}
\usetikzlibrary{matrix}
\usepackage{color}
\usepackage{graphicx,amssymb,amsfonts,amsmath,amssymb,amscd,amstext, mathrsfs}
\usepackage{graphicx} 
\usepackage{dsfont}
\usepackage{xcolor}
\usepackage{amsmath}
\usepackage{empheq}
\usepackage{cite}
\usepackage{bm}

\numberwithin{equation}{section}

\newlength\dlf

\def\cO{\mathcal{O}}

\newcommand{\bw}{\begin{widetext}}
\newcommand{\ew}{\end{widetext}}
\newcommand{\bea}{\begin{eqnarray}}
\newcommand{\eea}{\end{eqnarray}}
\newcommand{\be}{\begin{equation}}
\newcommand{\ee}{\end{equation}}
\newcommand{\nn}{\nonumber}

\renewcommand{\bar}[1]{\overline{#1}}
\renewcommand{\tilde}[1]{\widetilde{#1}}

\newcommand{\<}{\langle}
\renewcommand{\>}{\rangle}
\newcommand{\norm}[1]{\left\lVert#1\right\rVert}

\renewcommand{\cal}{\mathcal}
\newcommand{\CO}{\mathcal{O}}

\newcommand{\CN}{\mathcal{N}}

\newcommand{\CF}{\mathcal{F}}

\newcommand{\CL}{\mathcal{L}}

\newcommand{\kvec}{ {\boldsymbol{k}} }

\newcommand{\p}{\partial}

\newcommand{\bk}{\boldsymbol{k}}

\DeclareFontShape{OT1}{cmr}{mx}{n}{<->cmr10}{}
\newcommand{\titlefont}{\fontseries{mx}\selectfont}

\def\frac#1#2{{#1\over #2}}

\newcommand{\newRho}{{\varrho}}

\usepackage{jheppub}

\begin{document}

\begin{titlepage}

\begin{flushright} 
\end{flushright}

\begin{center} 

\vspace{0.35cm}

{\fontsize{21.5pt}{0pt}{\titlefont 
Form Factors and Spectral Densities  from Lightcone Conformal Truncation
}}

\vspace{1.6cm}  

{{Hongbin Chen$^1$, A. Liam Fitzpatrick$^1$,  Denis Karateev$^{2}$}}

\vspace{1cm} 

{{\it
$^1$Department of Physics, Boston University, 
Boston, MA  02215, USA
\\
\vspace{0.1cm}
$^2$Philippe Meyer Institute, Physics Department \'Ecole Normale Sup\'erieure (ENS), Universit\'e PSL24 rue Lhomond, F-75231 Paris, France
}}\\
\end{center}
\vspace{1.5cm}

{\noindent 
We use the method of Lightcone Conformal Truncation (LCT) to obtain form factors and spectral densities of local operators $\CO$ in $\phi^4$ theory in two dimensions.  We show how to use the Hamiltonian eigenstates from LCT to obtain form factors
that are matrix elements of a local operator $\CO$ between single-particle bra and ket states, and we develop methods that significantly reduce errors resulting from the finite truncation of the Hilbert space.  We extrapolate these form factors as a function of momentum to the regime where, by crossing symmetry, they are form factors of $\CO$ between the vacuum and 
a two-particle asymptotic scattering state.  We also compute the momentum-space time-ordered two-point functions of local operators in LCT.  These converge quickly at momenta away from branch cuts, allowing us to indirectly  obtain the time-ordered correlator and the spectral density at the branch cuts.  We focus on the case where the local operator $\CO$ is the trace $\Theta$ of the stress tensor.
}

\end{titlepage}

\tableofcontents

\section{Introduction} 

Hamiltonian Truncation methods are a powerful approach to studying Quantum Field Theory (QFT) at strong coupling. Their basic strategy is to numerically diagonalize the Hamiltonian restricted to a finite dimensional subspace of the full Hilbert space.  The resulting energy eigenvalues and eigenvectors provide a wealth of information about the theory at strong coupling, but it is not always straightforward to assemble this output into the dynamical observables of interest.  

In this paper, we will focus on two particular types of observables involving local operators, namely spectral densities of two-point functions, and two-particle form factors.  We will use Lightcone Conformal Truncation (LCT), which is a specific Hamiltonian truncation method \cite{Katz:2016hxp,Anand:2020gnn}. Spectral densities, through the K\"allen-Lehmann spectral decomposition, can be used to easily reconstruct two-point correlators in position space or momentum space, in Euclidean or Lorentzian signature.  They describe the inclusive scattering cross-sections that result when one weakly couples the theory to an external probe. However, in a truncation framework, they suffer from the fact that one must necessarily perform some type of discretization of the multi-particle continuum when truncating the Hilbert space. Such discretizations are at odds with the typically smooth behavior of spectral functions on the multi-particle continuum in infinite volume. In the cases we consider, this discrepancy will be particularly offensive: the truncated spectral density is a sum over delta functions, but the continuum limit is supposed to be a smooth function in the multi-particle regime. One of our goals in this work will be to improve the spectral density to rectify this issue.  

The other type of observable we consider, the form factors, have their own scars from the truncation of the Hilbert space.   There are two kinds of form factors we consider, related to each other by crossing symmetry.  The first kind of form factor is a matrix element of a local operator $\CO$ between single-particle states, i.e. $\mathcal{F}_{1,1}^\CO \equiv \< m, p | \CO(0) | m, p'\>$.  The second is a matrix element of $\CO$ between the vacuum and a two-particle asymptotic scattering state, i.e. $\mathcal{F}_{2,0}^\CO \equiv \< {\rm vac} | \CO(0) | m, p; m, p'\>$.  The former can be computed directly in LCT, though as we will see, the act of truncation ruins some of the important analytic properties of the function $\mathcal{F}_{1,1}^\CO$ and one of our goals will be to undo this damage.  The second cannot be computed directly, since multi-particle asymptotic scattering states are not generally eigenstates of the Hamiltonian.  Instead, we will aim to improve the behavior of $\mathcal{F}_{1,1}^\CO$ enough to obtain $\mathcal{F}_{2,0}^\CO$ from it by analytic continuation.

We choose to specifically focus on the $\phi^4$ model in 2d:
\begin{equation}
\label{eq:UV_theory}
S_{UV}=\int d^2x\left(-\frac{1}{2} (\partial\phi)^2- \frac{1}{2} m_0^2 \phi^2 - \frac{\lambda}{4!} \phi^4
\right).
\end{equation}
The main reason for this choice is that LCT has been developed furthest for this model, and so this is the model where we will be able to obtain the highest quality truncation data; we expect the approach in this paper to apply more generally.

We will define all operators in the Lagrangian to be normal-ordered, which removes all divergences in the theory.  The theory  depends only on the dimensionless quartic coupling $\bar \lambda$ defined as\footnote{Notice the mismatch of notation with the work \cite{Anand:2020gnn}. One has $\bar \lambda_\text{here} = 4\pi \bar \lambda_\text{there}$.}
\begin{equation}
\label{eq:lambda_bar}
\bar \lambda \equiv m_0^{-2}\lambda.
\end{equation}
  We will work in lightcone quantization, which shifts the value of the bare mass-squared $m_0^2$ relative to equal-time quantization \cite{Burkardt,Burkardt2,Fitzpatrick:2018xlz}. 
The theory is in the unbroken phase in  range  $0 \le \bar{\lambda} \le \bar{\lambda}_* $, where $\bar{\lambda}_*\approx 23.1$  \cite{Anand:2017yij}.
At the critical value $\bar \lambda_*$, the theory flows in the IR to 2d Ising model.\footnote{See e.g. \cite{Chabysheva:2015ynr,Burkardt:2016ffk,Elliott:2014fsa,Chabysheva:2016ehd,Anand:2017yij} for various recent works applying Hamiltonian truncation methods to 2d $\phi^4$ theory, and in particular \cite{Bajnok:2015bgw} which extracted the scattering phase shifts in the broken phase of the theory with the aid of powerful equal-time renormalization methods from \cite{Rychkov:2014eea,Rychkov:2015vap,Elias-Miro:2017xxf,Elias-Miro:2017tup,Elias-Miro:2015bqk}. }

\paragraph{Summary of Main Results}

In this paper, we use the LCT method to compute the form factor of the trace of the stress tensor  
and the  spectral density of the trace of the stress tensor $\rho_\Theta(s)$.  One of our main results is a procedure that leads to significant improvements in the spectral densities of local operators, compared to previous LCT work.  In Fig. \ref{fig:SDNonPerturbative}, we show our improved spectral density in $\phi^4$ theory for a range of couplings over the entire unbroken phase region $0 \le \bar{\lambda} \le 23.1$; for $\bar\lambda\ll4\pi$, our truncation results match the perturbative ones very well.   As mentioned above, the ``raw'' truncation result for the spectral density is a sum over $\delta$ functions, since the truncated spectrum is discrete, so any smooth result for the spectral density is already an improvement.  But our procedure also significantly improves integrals of spectral densities, as the $C$-function plotted in Fig. \ref{fig:SDIntegrated} shows.  Our method uses the fact that the time-ordered correlator converges fairly quickly away from poles and branch cuts, and the derivatives of the time-ordered correlator in the convergent region can be used to reconstruct its behavior near the branch cuts.

We also show how form factors  $\mathcal{F}_{1,1}^\CO$  for any local operator $\CO$ at physical external values of the particle momenta can be obtained directly from the one-particle eigenvectors of the Hamiltonian in LCT.  The ``raw'' result in this case is obtained by applying a simple formula (\ref{eq:TruncFormFactorBasic}) to the LCT data. We then use this raw result as a starting point for various improvements that significantly reduce truncation effects. The final results are given in figures \ref{fig:SDNonPerturbative} and \ref{fig:FFplot} for various values of $\bar\lambda$. Again, for $\bar\lambda\ll4\pi$, the match to perturbation theory is quite good.  The form factor $\mathcal{F}_{1,1}^\CO$ at physical external momenta is equivalent by crossing symmetry to $\mathcal{F}_{2,0}^\CO(s)$ at $s<0$.  By contrast with $s<0$, there is no direct way of computing $\mathcal{F}_{2,0}^\CO(s)$ with $s>0$ using the LCT method.  In principle, it can be obtained by analytic continuation  from $s<0$ to $s>0$, and in this paper we attempt to perform this extrapolation with reasonable results shown in Fig. \ref{fig:ffvsSD} and \ref{fig:ffpositives}. As one can see in Fig.  \ref{fig:ffvsSD}, at strong coupling this extrapolation produces results that do not quite satisfy unitarity consistency conditions in the elastic regime $4m^2 < s < 16m^2$.   In a companion paper \cite{truncboot}, we propose to obtain the form factors, as well as the S-matrix, in this elastic regime via the S-matrix/form factor bootstrap program with the observables computed in this paper as the input.

\paragraph{Outline of the paper}
In section \ref{sec:review_LCT}, we give a brief review of the LCT method. In section \ref{sec:SDandPade}, we first discuss how one can compute spectral densities and time-ordered correlators directly using the LCT eigenvalues and eigenvectors. We then propose to use a two-point Pad\'e approximant of the time-ordered correlators to improve the result for the spectral densities. In section \ref{sec:LCT_FF}, we compute the two-particle form factor of the trace of the stress tensor at $s\le0$, and discuss how one can obtain more accurate result by using various tricks. We then make a first attempt to go to the $s>0$ regime directly by analytic continuation. We end with a brief discussion in section \ref{sec:discussion}.  A number of technical details of our approach are relegated to appendices. Additionally, in appendix \ref{app:perturbativeLCT},  we  check the results of our method in the limit of perturbative coupling $\lambda$, as well as at large $N$ in the $O(N)$ generalization of the model, where we find excellent agreement with the standard perturbative Feynman diagram approach.

\section{Lightning Review of LCT} 
\label{sec:review_LCT}
 
 The LCT approach 
 works with the Hamiltonian in lightcone quantization, so states are defined on a null plane and the Hamiltonian $P_+$ evolves them forward in lightcone time $x^+$.  The relation between the lighcone and cartesian coordinates is given by
\begin{equation}
x^{\pm} \equiv \frac{x^0\pm x^1}{\sqrt{2}}.
\end{equation} 
The advantage of lightcone quantization is that the vacuum does not mix with other states in the theory, so there are no ``vacuum bubble'' diagrams and one may formally take the limit of infinite volume from the outset. Additionally, LCT decomposes the Hamiltonian as $H=H_0+V$, where $H_0$ is the Hamiltonian for a UV CFT, and $V$ is one or more relevant deformations.  The truncation subspace is then taken to be all states that sit inside a representation of the conformal group with quadratic Casimir below some cut-off.\footnote{Discrete Lightcone Quantization (DLCQ) \cite{Pauli:1985ps}  uses a different basis constructed by first compactifying in the lightlike direction. See \cite{Harindranath:1987db,Harindranath:1988zt} for an early application to 2d $\phi^4$ theory.}  Roughly speaking, one keeps all states created by operators with dimension below some cutoff $\Delta_{\rm max}$.  More precisely, the basis states are of the form 
 \be
 \label{eq:states}
 |\CO, p\> \equiv \frac{1}{N_{\CO}} \int d^d x e^{-i p \cdot x} \CO(x) | {\rm vac} \>,
 \ee
 where $\CO$ is a primary operator of the UV CFT with dimension $\le \Delta_{\rm max}$, and $N_\CO$ is a normalization constant.  
 
 The practical advantage of this basis is that conformal symmetry highly constrains the correlation functions of such operators, which in turn increases the efficiency of computing the matrix elements of the Hamiltonian.  We will see shortly that it also greatly aids the computation of form factors of local UV operators.
 
The mass-squared operator
\begin{equation}
M^2 \equiv -P^2 = 2 P_+ P_- - P_\perp^2
\end{equation}
depends linearly on the lightcone Hamiltonian $P_+$. We are free to work in a momentum frame where $P_\perp=0$, and $P_-=p_-$ is some value of our choosing.  When we diagonalize $P_+$, we can label the eigenvectors by their $M^2$ and $P_-$ eigenvalues $\mu^2$ and $p_-$, and write them as a sum over our basis states:
\be
| \mu_i^2, p_- \> = \sum_{\CO_j}  C_{\CO_j}^{\mu_i^2} | \CO_j , p\>\quad \text{with } \mu_i^2=2p_{+i}p_{-}. 
\label{eq:EstatesInLCTBasis}
\ee
Because the quantization surface $x^+=0$ preserves boosts in the $x^1$ direction, the matrix elements of $M^2= 2 p_- P_+$ are invariant under such boosts.  A very useful consequence is that diagonalizing $P_+$ at one value of $p_-$ effectively gives us the eigenvectors for all values of $p_-$.

Having diagonalized the truncated Hamiltonian $P_+$ (or equivalently $M^2$), one can compute several observables from the eigenvalues and eigenvectors. In this work, we focus on two such observables: the spectral density, which we address in section \ref{sec:SDandPade}, and the two-particle form factor, which we address in section \ref{sec:LCT_FF}. For the definitions of the observables and conventions used in this paper, see appendix \ref{app:definitions}.

\section{Spectral Densities}
\label{sec:SDandPade}

The most natural quantity to compute in the LCT framework is the spectral density $\rho_{\CO}$ of local operators.  In terms of the eigenvectors of $M^2$, for a scalar operator $\CO$ it is simply 
\be
\rho_\CO(s) = \sum_i | \<{\rm vac}| \CO(0) | \mu_i^2, p_-\> |^2 \delta(s- \mu_i^2),
\label{eq:SDdefn}
\ee
where the sum in $i$ is a sum over states in a fixed momentum frame. 
From \eqref{eq:SDdefn} one can easily obtain the $\CO$ two-point function in Lorentzian or Euclidean signature, in position space or momentum space.  For instance, the Fourier transform of the time-ordered correlator is (with $s=-p^2$)
\begin{align}
\mathbf{ \Delta}_{\CO}(p) &\equiv \int d^d x e^{i p \cdot x} \<{\rm vac}|  \CO(x) \CO(0) |{\rm vac} \>_T\nn\\
& = \int_0^\infty d\mu^2 \rho_{\CO}(\mu^2) \frac{i}{s-\mu^2 + i \epsilon} = \sum_i \frac{i | \<{\rm vac}| \CO(0) | \mu_i^2, p_-\> |^2}{s - \mu_i^2+i \epsilon} .
\label{eq:TOCandSD}
\end{align} 
Due to \eqref{eq:states} and \eqref{eq:EstatesInLCTBasis}, the matrix element entering in \eqref{eq:SDdefn} and \eqref{eq:TOCandSD} can be simply written as linear combinations of Fourier transforms of CFT two-point functions of local operators as
\begin{equation}
\<{\rm vac}| \CO(0) | \mu_i^2, p_-\> = 
\sum_{\CO_j}  C_{\CO_j}^{\mu_i^2}\frac{1}{N_{\CO_j}}
\int d^d x e^{-i p \cdot x}
\<{\rm vac}| \CO(0)  \CO_j(x) | {\rm vac} \>. 
\end{equation}
When $\CO$ is a CFT operator in the UV CFT, the two-point function on the RHS is the usual CFT two-point function in the Lorentzian signature, and without loss of generality we can choose an operator basis so that it is only non-vanishing if $\CO_j=\CO$. 
Its momentum space expression\footnote{For 2d, it is simply given by 
\begin{equation}
	\int d^{2} x e^{-i p \cdot x}\langle{\rm vac}|\mathcal{O}(0)\mathcal{O}(x)|{\rm vac} \rangle=\frac{4\pi^2 p_+^{2 h-1}p_-^{2\bar h-1}}{\Gamma(2 h)\Gamma(2\bar h)} \Theta(p_+)\Theta(p_-)
\end{equation}
where $h$ and $\bar h$ are the conformal dimensions of $\CO$, and the position space two-point function is normalized to have norm equal to 1 in the above formula. For holomorphic operators $h=0$, the limit $h \rightarrow 0$ of the above formula produces a $\delta(p_+)$ function.} can be found in \cite{Anand:2019lkt}.  
If $\CO$ is an operator in our basis and we choose the normalization coefficients $N_\CO$ so that $\< \CO, p | \CO, p'\> = 2 p_- (2\pi) \delta(p_- - p_-')$, then the overlap is simply\footnote{See \cite{Anand:2020gnn}, eq (4.81).}
\begin{equation}
\<{\rm vac}| \CO(0) | \mu_i^2, p_-\> = C_\CO^{\mu_i^2} 2p_- N_\CO.
\end{equation}

In Hamiltonian truncation methods, the Hilbert space is truncated to a finite-dimensional subspace.  As a consequence, the eigenvalue spectrum of the Hamiltonian typically is discrete and therefore the spectral densities computed using equation (\ref{eq:SDdefn}) are sums over $\delta$ functions.\footnote{One way around this statement is to use a truncated basis that is different in each momentum frame, so that a continuous spectrum of eigenvalues arises from the continuum of momentum frame choices.  See e.g. \cite{Delacretaz:2018xbn} for an example of such an approach in the large $N$ limit of the 3d $\CO(N)$ model and Chern-Simons theories.} However, the exact spectral densities have continuous contributions from multi-particle states.  Obviously a sum over $\delta$ functions can at best reproduce a continuous function in some distributional sense but not in an absolute sense.  So for instance, the integrated spectral density (or more generally, the spectral density integrated against any smooth kernel) may exhibit absolute convergence to  its continuum limit.  

If we want precise results for the spectral densities from Hamiltonian truncation, we therefore seem to have two possible options.  Either, we formulate all applications of the spectral densities in terms of weighted integrals thereof, or we process the spectral densities in some way to make them continuous.  We think that both of these approaches are worth pursuing.  Here, we will focus on the latter approach, in an attempt to construct a smooth spectral density.  Our motivation for focusing on this direction is that if we can significantly improve the convergence of the spectral density itself (from convergence in a distributional sense to convergence in an absolute sense), it seems likely that the integrated moments of the spectral density will also be improved.\footnote{And in fact we check explicitly in Fig. \ref{fig:SDIntegrated} that we improve the integrated spectral density by our methods.} However, the method we introduce will involve some guesswork, and it may be the case that integrated spectral densities would allow greater mathematical rigor. 

\subsection{Pad\'e Approximation}
\label{sec:SD_Pade}

The basic idea is to focus first not on the spectral density itself, but rather on the time-ordered two-point function. If we can obtain an accurate result for the time-ordered two-point function, then the spectral density is simply its real part
\begin{equation}
	2\pi\theta(p^0)\rho_\CO(-p^2) = 2\,\text{Re}
\int d^dx e^{-i p\cdot x}
\< {\rm vac} |\CO(x)\CO(0)|{\rm vac}\>_T.
\end{equation}
 After factoring out an overall momentum-conserving $\delta$ function, the time-ordered two-point function is an analytic function of $s$ away from poles and cuts corresponding to physical states.  Therefore, at values of $s$ in the complex plane away from the branch cut, we can expect much better convergence of the time-ordered correlator.  The trick is to then use this convergence of the function away from the branch cut to reconstruct its behavior on the branch cut. 

\begin{figure}[!t]
\begin{center}
\includegraphics[width=0.48\textwidth]{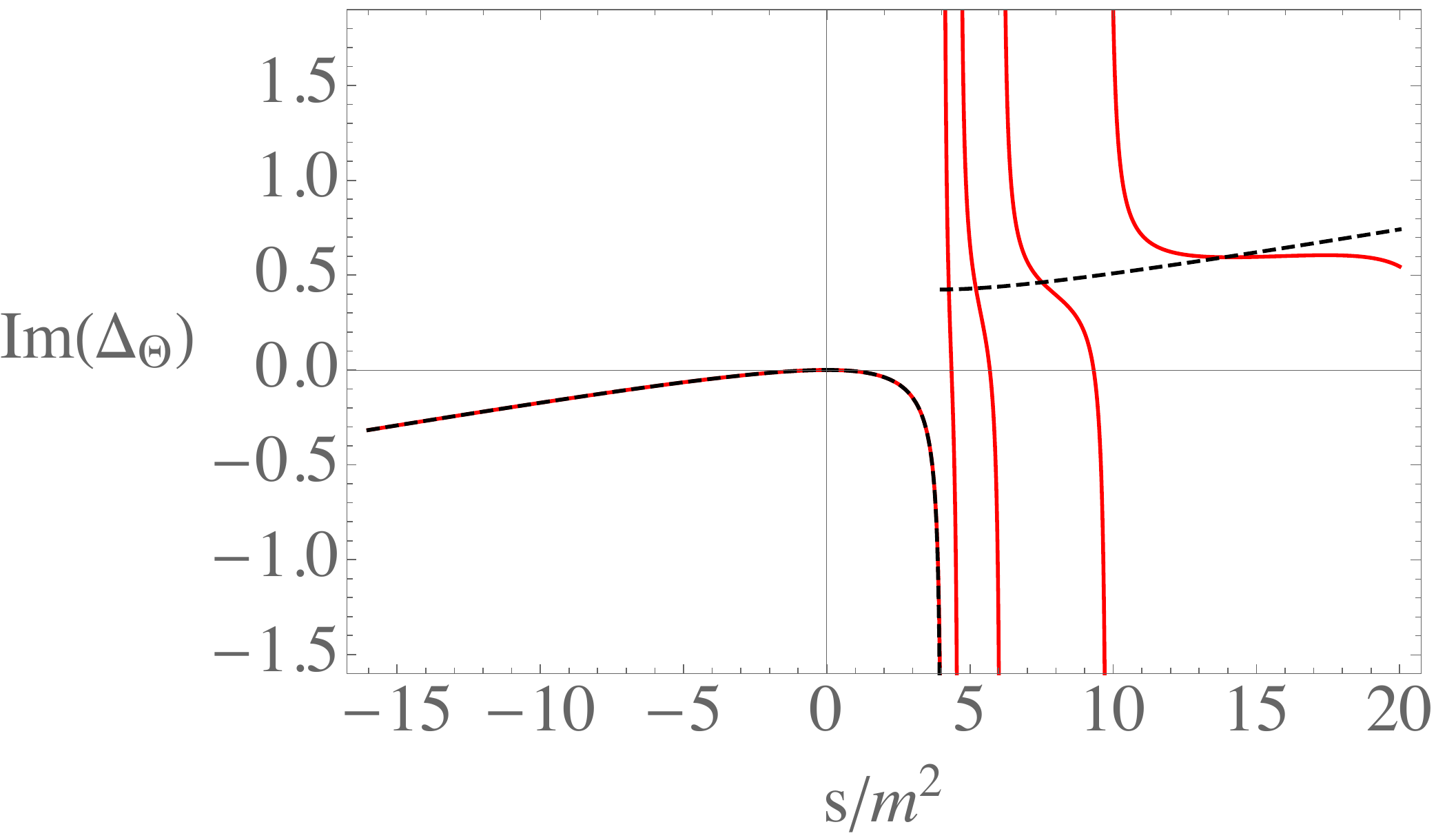}
\caption{Imaginary part of the time-ordered correlator $\mathbf{\Delta}_\Theta(s)$ from LCT using the spectral representation (\ref{eq:TOCandSD}) at $\Delta_{\rm max}=12$ ({\it red, solid}) compared to the exact result ({\it black, dashed}) in the free scalar theory (\ref{eq:free2pt}).  Because the truncated Hamiltonian has a discrete spectrum, the resulting time-ordered correlator has a series of discrete poles at $s>4m^2$, whereas the continuum limit should give  a smooth function as shown.  The real part from LCT is not shown because it is a sum of $\delta$ functions. }
\label{fig:SDFree}
\end{center}
\end{figure}

\begin{figure}[t]
\begin{center}
\includegraphics[width=0.96\textwidth]{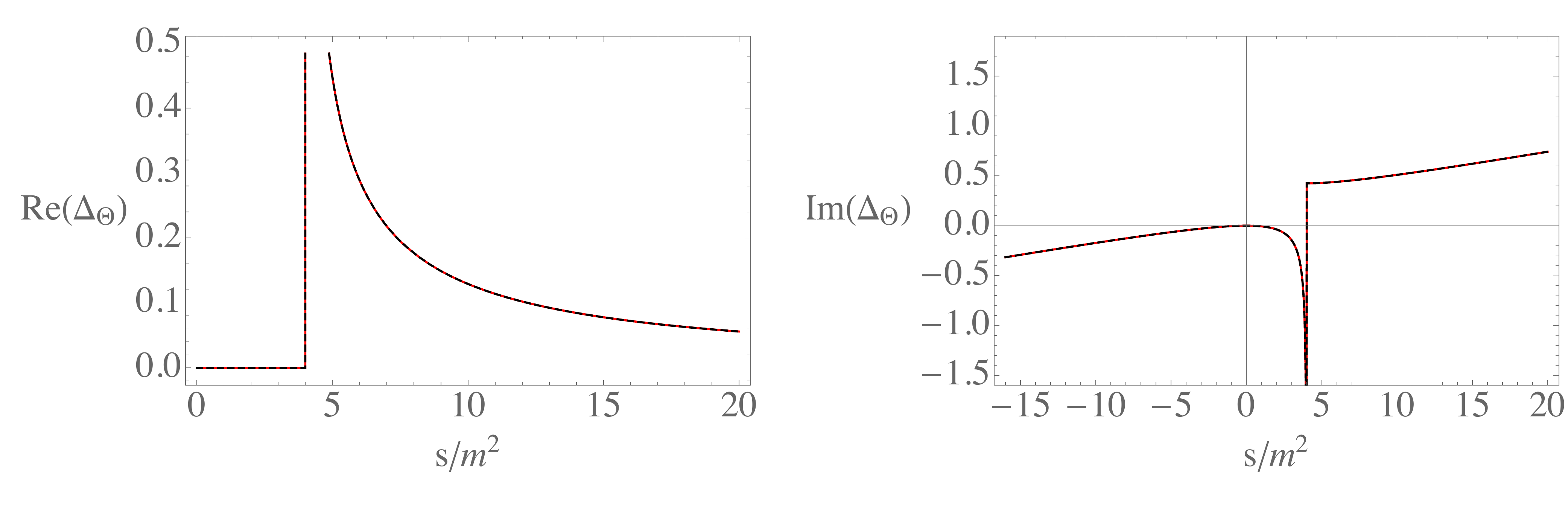}
\caption{Real ({\it left}) and imaginary ({\it right}) parts of the time-ordered correlator $\mathbf{\Delta}_\Theta(s)$ in the free scalar theory, comparing the exact result ({\it black, dashed}) to the LCT result at $\Delta_{\rm max}=12$ ({\it red, solid}) after using Pad\'e approximants as described in the text.  }
\label{fig:SDFree2}
\end{center}
\end{figure}

As a simple example, consider the spectral density $\rho_{\Theta}$ of the stress tensor in the 2d free theory, $\bar\lambda=0$.  The exact result, from a one-loop computation, is that 
\be\label{eq:free2pt}
\mathbf{\Delta}_\Theta(p) \equiv \int d^2 x e^{-i p \cdot x} \<{\rm vac} | \Theta(x) \Theta(0) | {\rm vac}\>_T = \frac{1}{2 \pi i} \left( \Delta(s) - \frac{s}{6} \right),
\ee
where $\Delta(s)$ is given in (\ref{eq:DeltaDef}).  In Fig. \ref{fig:SDFree}, we show a comparison of the imaginary part 
of this exact time-ordered correlator above against the result from LCT with $\Delta_{\rm max}=12$.  The truncation result at $s>4m^2$ has poles at the eigenvalues of the truncated Hamiltonian, and is a very poor approximation to the true correlator (the real part is a sum over $\delta$ functions and cannot even be plotted).  However, note that at $s<4m^2$, LCT gives a very good approximation, even with this relatively small value of $\Delta_{\rm max}$.  To take advantage of this faster convergence away from $s>4m^2$, we can compute the Taylor coefficients in $s$ around some point in the complex plane  and use these series coefficients to reconstruct a function with smooth real and imaginary parts near the branch cut. This can be done via using the Pad\'e approximation.  Note that the time-ordered two-point function computed in LCT using the spectral representation (\ref{eq:TOCandSD}) is a symbolic function of $s$, and therefore derivatives around any point may be taken symbolically rather than numerically.   As usual, to improve convergence, it helps to use the variable $\newRho$ defined via
\begin{equation}
\label{eq:def_rho_1}
s = \frac{ 16 m^2 \newRho}{(1+ \newRho)^2}.
\end{equation}
This maps the cut plane to the unit disk. As an example, in Fig. \ref{fig:SDFree2}, we show that taking the order $(5,5)$ Pad\'e approximant\footnote{Following convention, we refer to the rational function of the form $\frac{\sum_{i=0}^n a_i z^i}{1+ \sum_{j=1}^m b_j z^j}$ with series coefficients matching those of $f(z)$ up to $\CO(z^{n+m})$ as the $(n,m)$ Pad\'e approximant of $f$ around $z=0$.}
around the point $\newRho=0$ produces a function whose real and imaginary parts are very good approximations to the time-ordered correlator even at $s>4m^2$. Although this method requires some care -- in particular, taking the order of the Pad\'e approximant too high or too low gives poor results, and it depends on which points in $s$ one chooses to expand around -- it can be a powerful way to improve the calculation of the spectral densities from Hamiltonian truncation.  

There are a couple additional improvements we can make to the method.  The first is that we can do a ``multi-point'' Pad\'e approximation where we fix the Taylor coefficients of the approximant around multiple points.  We will limit ourselves to two points, $s=0$ and $s=\infty$.  The advantage of $s=\infty$ (which is $\newRho=-1$) is that the large $s$ limit of the theory is free and therefore controlled by perturbation theory, even when $\bar\lambda$ is large.  We can in fact fix the first two powers of $1/s$ exactly with very little work.  For the stress tensor one has\footnote{It is important that the bare mass-squared $m_0^2$ that appears in (\ref{eq:TOCsubleading}) is the bare mass in {\it lightcone} quantization, which differs from the bare mass in equal-time quantization due to zero modes \cite{Burkardt,Burkardt2,Fitzpatrick:2018xlz}.  It is interesting and perhaps surprising that the large $s$ expansion of the $\Theta$ two-point function, which is a simple physical observable, should be so directly related to the lightcone bare mass.}  
\be
\label{eq:TOCsubleading}
\mathbf{\Delta}_{\Theta}(p) = s^2 \mathbf{\Delta}_{T_{--}}(p)=  \frac{is^2}{12 \pi } \left( \frac{1}{s}  + \frac{6m_0^2 + \frac{3 \lambda}{4 \pi} }{s^2} +\dots \right) .
\ee
We provide the detailed derive of this result in appendix \ref{app:large_energy}.
We can use this knowledge to fix the coefficients in an expansion in $\newRho$ about $\newRho \sim - 1$ up to $\CO((1+\newRho)^5)$, since the next term $1/s^3$ will only contribute at order $\CO((1+\newRho)^6)$.

\begin{figure}   
 \centering
 \includegraphics[height=4.5cm]{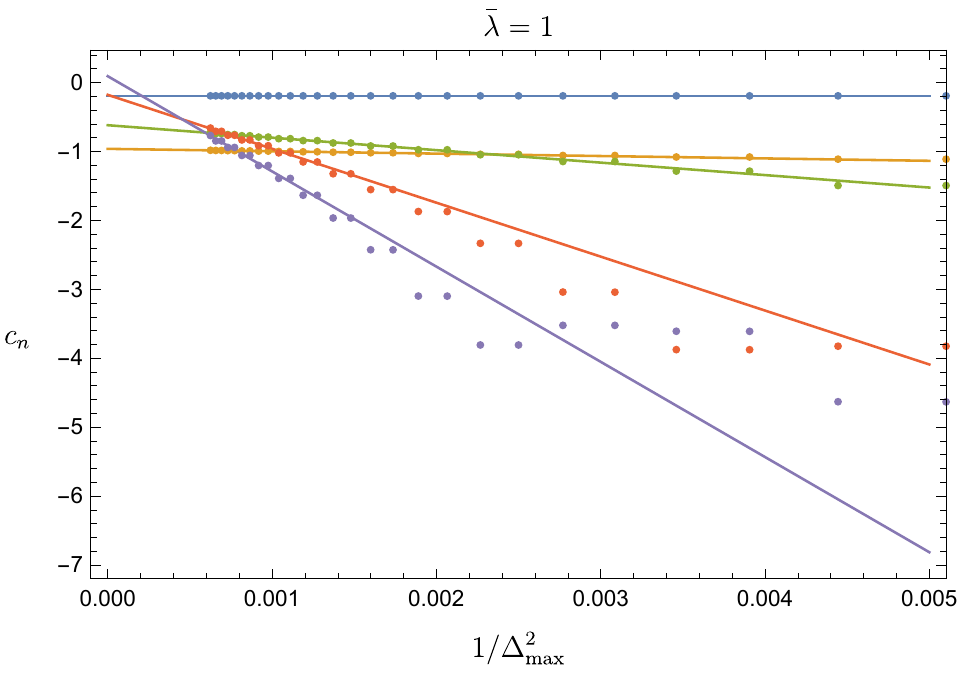} 
 \quad
  \includegraphics[height=4.5cm]{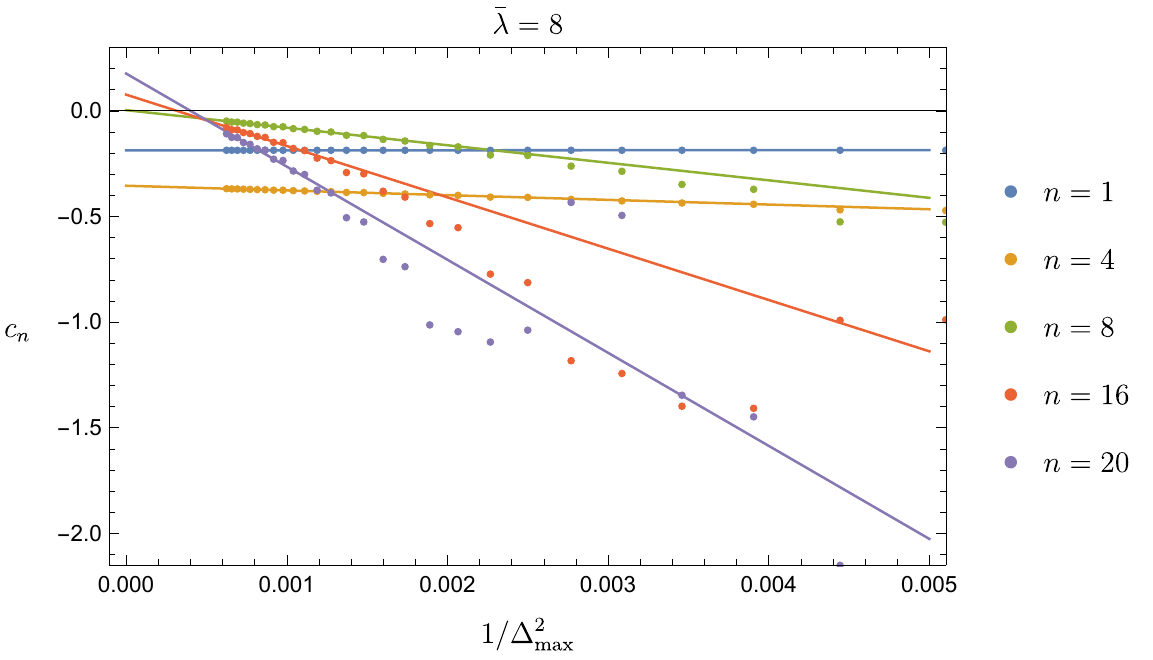} 
        
 \includegraphics[height=4.5cm]{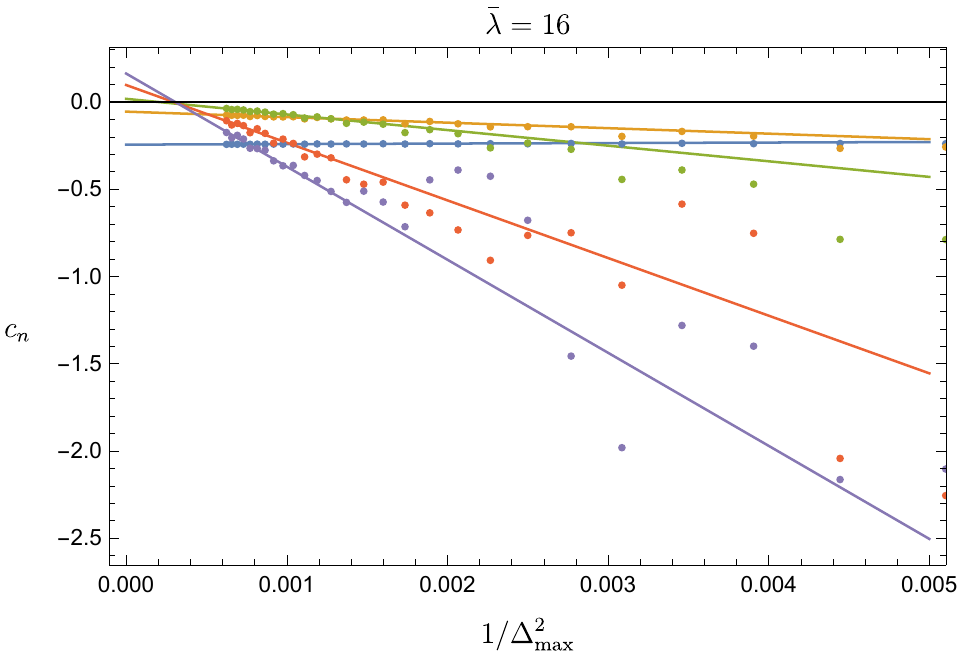} 
 \quad
\includegraphics[height=4.5cm]{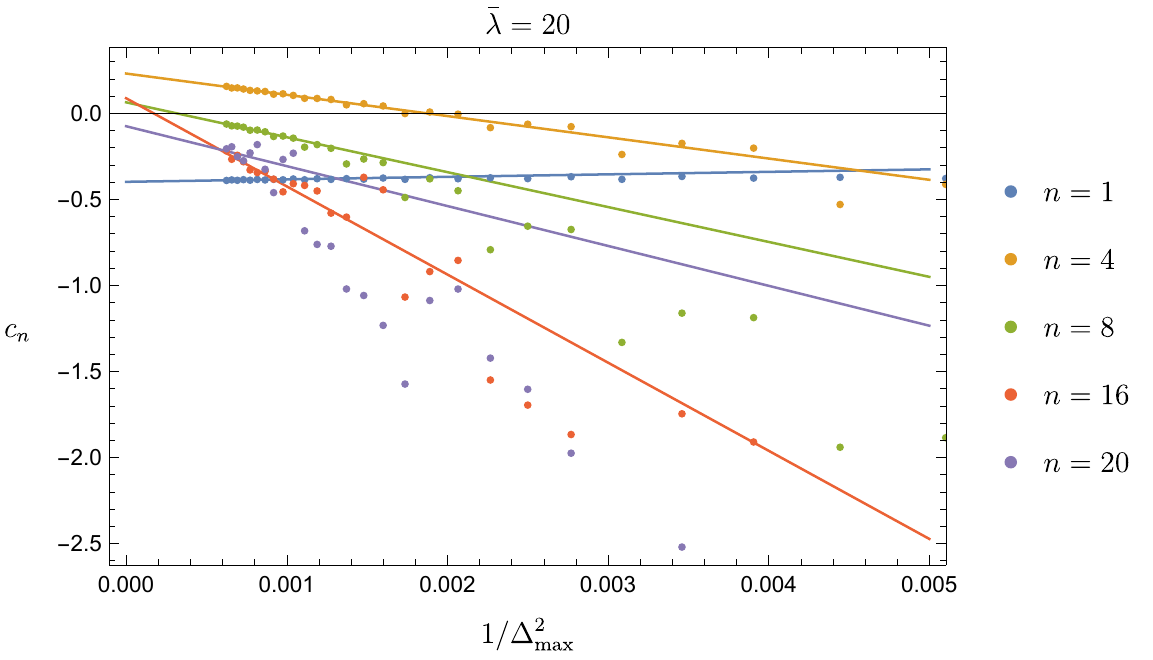} 
 \caption{Convergence of the Taylor coefficients $c_n$ of $\mathbf{\Delta}_\Theta(\newRho(s))=\sum_{n=0}^\infty c_n \newRho^n$ at $\newRho=0$. For a given $\bar\lambda$, we computed the coefficients $c_n$ for each $\Delta_\text{max}$ up to $\Delta_\text{max}=40$, and then extrapolated them to $\Delta_\text{max}=\infty$ by fitting them as a function of $x=1/\Delta_\text{max}$. The function we used to fit these coefficients is $a+ b x^2$, and the solid lines are the results of the fits. }
 \label{fig:ConvCoeSD}
\end{figure}

One final advantage of computing Taylor coefficients in $s$ or $\newRho$ is that each coefficient can be computed for any value of $\Delta_{\rm max}$, and then we can attempt to extrapolate them to $\Delta_{\rm max} = \infty$.  In practice, we have found that this convergence is fastest if we fix the mass gap to be the same for each value of $\Delta_{\rm max}$; this requires dialing the coupling $\bar\lambda$ as a function of $\Delta_{\rm max}$ to keep the gap fixed. In figure \ref{fig:ConvCoeSD},  we show the convergence for some of the Taylor coefficients of the time-ordered correlator $\mathbf{\Delta}_\Theta(\newRho)=\sum_{n=0}^\infty c_n \newRho^n$ at $\newRho=0$. In general, for a given $n$, the convergence slows down as we increase $\bar\lambda$, and for a given $\bar \lambda$, the convergence is becoming worse for larger $n$.

\subsection{Final Results for the Spectral Density}
\label{sec:SD_final_results}
\begin{figure}[!t]
\begin{center}
\includegraphics[width=0.96\textwidth]{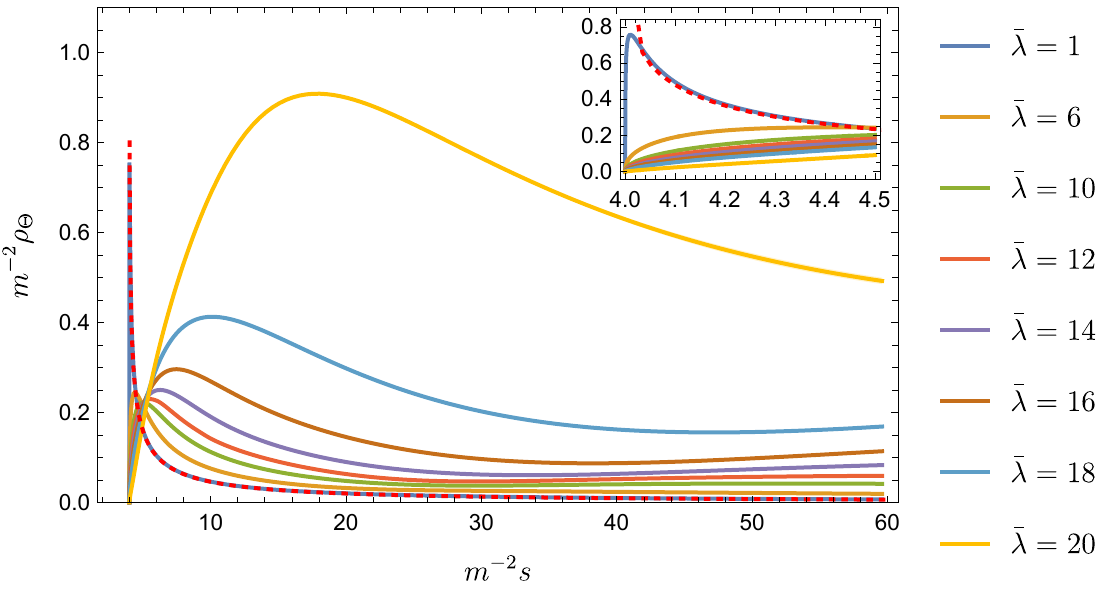}
\caption{Spectral densities of the trace of the stress tensor $\rho_\Theta$ in the $\phi^4$ model computed for various values of $\bar\lambda$ from LCT (with the inset showing more details near $s=4m^2$). These plots are obtained by taking the imaginary part of the two-point Pad\'e approximant of the trace of the stress tensor two-point function as described in the main text. As a comparison, we also plotted the perturbative spectral density for $\bar\lambda=1$ (red dotted line) computed via $\left|\mathcal{F}_{2,0}^{\Theta}(s)\right|^{2}/(2 \pi \mathcal{N}_{2})$, where $\mathcal{F}_{2,0}^{\Theta}(s)$ is the two-loop form factor given in equation (\ref{eq:FF_pert}). One can see that it agrees with the LCT non-perturbative $\bar\lambda=1$ result very well at large $s$ as expected, since $\bar\lambda=1\ll4\pi$ is in the perturbative regime.  This provides a consistency check for our procedure for computing the non-perturbative spectral density from LCT.
However, one can also see the difference near $s=4m^2$, where perturbative theory breaks down. Especially, the perturbative result has a singularity at $s=4m^2$, while the non-perturbative LCT result is regular there. 
 }
\label{fig:SDNonPerturbative}
\end{center}
\end{figure}

Let us summarize our approach for computing the spectral densities. We use LCT to compute time-order two-point function $\mathbf{ \Delta}_{\Theta}$ at different values of $\Delta_\text{max}$ up to $\Delta_\text{max}=40$ as a symbolic function of $s$, and Taylor expand around $s=0$ (equivalently, $\newRho=0$). We then extrapolate these Taylor coefficients to $\Delta_\text{max}=\infty$. Combining with the several Taylor coefficients from equation (\ref{eq:TOCsubleading}) at $\newRho=-1$, we perform a two-point diagonal Pad\'e approximation to obtain a rational function expression for $\mathbf{ \Delta}_{\Theta}$ in $\newRho$ that behaviors nicely at $s>4m^2$. Taking the imaginary part of this rational expression then gives us the spectral density $\rho_\Theta$. We show the final result we got for the spectral densities for various values of $\bar\lambda$ in figure \ref{fig:SDNonPerturbative}. The mass gaps in unit of $m_0$ for various values of $\bar \lambda$ are given in Table \ref{tab:gaps} for reference.

\begin{table}[t!]
\begin{small}
\begin{center}
\begin{tabular}{|c|c|c|c|c|c|c|c|c|c|c|}
\hline 
$\overline{\lambda}$ & 1 & 3 & 6 & 8 & 10 & 12 & 14 & 16 & 18 & 20\tabularnewline
\hline 
$m/m_0$ & 0.9988 & 0.9901 & 0.9637 & 0.9372 & 0.9025 & 0.8579 & 0.8000 & 0.7236 & 0.6186 & 0.4629\tabularnewline
\hline 
\end{tabular}
\caption{Mass gaps as a function of dimensionless coupling $\bar{\lambda}$.}
\label{tab:gaps}
\end{center}
\end{small}
\end{table}

There are various sources of uncertainty in our final result for the spectral densities. First, we chose the function $a+b x^2$ with $x=1/\Delta_\text{max}$ to extrapolate the coefficients to $\Delta_\text{max}=\infty$, which seems to work well, but we do not know if this is the correct asymptotic rate of convergence, and also the best fit parameters depend on the number  $p$ of points used in the fit. Second, when performing the two-point Pad\'e approximation, we also need to choose  what order of the Pad\'e approximant to use. We used a simple procedure to determine these parameters: we scan over some reasonable range of  $p$, and then obtain the result for different orders of the Pad\'e approximant.  We then choose $p$ such that there exist three or four consecutive orders of the Pad\'e approximant   that give almost the same result for $s\in[4m^2, 100m^2]$. For example, for $\bar\lambda\le10$, the results we obtained this way for the spectral density are very similar for Pad\'e approximants of orders between $(14,14)$ and $(17,17)$ (both diagonal and non-diagonal)\footnote{Note that we use six Taylor coefficients of the time-order two-point function $\mathbf{\Delta_{\Theta}}$ at $\newRho=-1$ as determined by equation (\ref{eq:TOCsubleading}) for the two-point Pad\'e approximant, so for order $(n,n)$, we will need $2n-6$ Taylor coefficients at $\newRho=0$.}. For larger $\bar\lambda$, we had to consider lower order Pad\'e approximants due to slower convergence of the higher order Taylor coefficients (as can been seen from the last plot in figure \ref{fig:ConvCoeSD}), e.g., for $\bar \lambda=20$, we looked at Pad\'e approximants of orders between $(8,8)$ and $(10,10)$. If we simply consider the uncertainty from the differences in these different orders of Pad\'e approximants, then it is $\CO(10^{-6})$ at $\bar\lambda=1$, and $\CO(10^{-2})$ at $\bar\lambda=20$ in the range $s\in [4m^2, 100m^2]$. 
 Of course, this is at most a lower bound of the actual uncertainty in the spectral densities we obtained.

\begin{figure}[t]
\begin{center}
\includegraphics[width=0.8\textwidth]{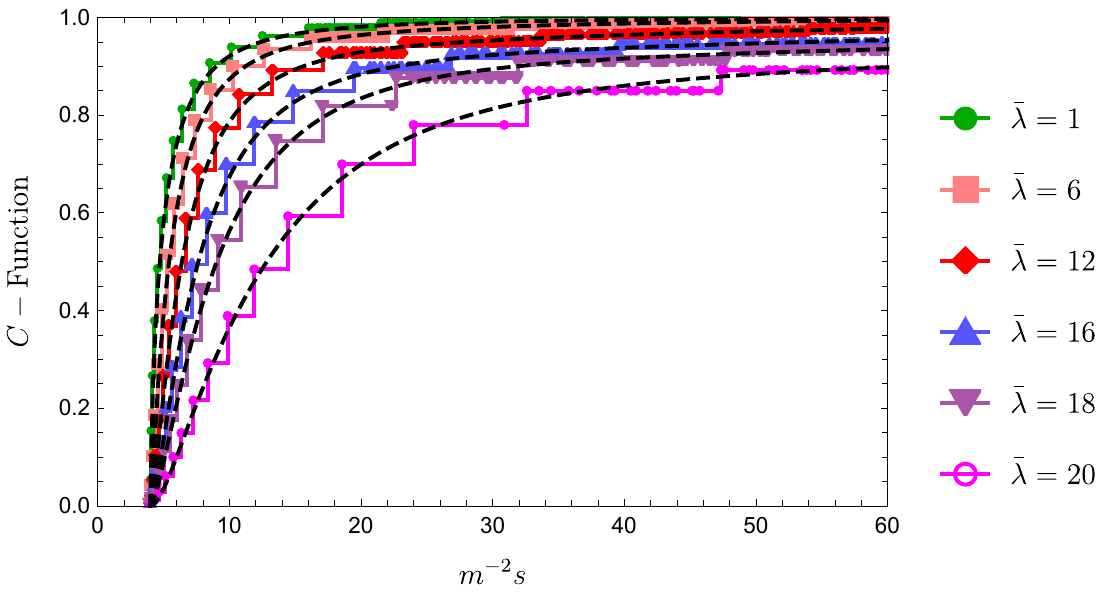}
\caption{The $C$-functions for various coupling constants. The piecewise continuous lines are computed from the truncation data with $\Delta_\text{max}=40$ directly (that is, by integrating the $\delta$ functions in equation (\ref{eq:SDdefn})), while the dashed lines are computed from integrating the Pad\'e approximant of the spectral density of $\Theta$ (i.e., real part of the Pad\'e approximant of the $\<\Theta\Theta\>$ two-point function). One can see that they agree with each other fairly well, but the Pad\'e approximant smooths out the unphysical steps in the raw truncation result.}
\label{fig:SDIntegrated}
\end{center}
\end{figure}

In figure \ref{fig:SDIntegrated}, we also show the $C$-function
\begin{equation}
	C(s)=12\pi \int_{4m^2}^s ds' \frac{\rho_\Theta(s')}{s^{'2}} 
\end{equation}
 computed by integrating the Pad\'e approximated expression of the spectral density, where one can see that  it is consistent with the result from simply integrating the $\delta$ function expression (\ref{eq:SDdefn}). This provides a consistency check of  our procedure for obtaining the spectral density.

\subsection{Pad\'e Approximation: general comments}

Finally, we end this section with some general comments about using Pad\'e approximants for  time-ordered correlators.  Because of their spectral representation, time-ordered correlators are  Stieltjes functions, defined as functions $f(x)$ on the cut plane $\mathbb{C}/\mathbb{R}_{x<0}$ of the form
\be
f(x) = \int_0^\infty dt \frac{\mu(t)}{1+x t},
\ee
for some non-negative measure $\mu(t)$ that decays sufficiently rapidly at $t \rightarrow \infty$ that the moments $\int_0^\infty  t^n \mu(t) dt$ exist for all positive integer $n$.  Clearly, $\mathbf{\Delta_{\CO}}(-s)$ for any operator $\CO$ is of this form, with $\mu(t)=t^{-1}  \rho_{\CO}(t^{-1})$. Convergence of the moment integrals at $t \sim \infty$ follows in a gapped theory  from the fact that $\rho(\mu^2)=0$ when $\mu$ is smaller than the gap of the theory; convergence at $t \sim 0$ follows for any operator $\CO$ in the UV CFT basis from the argument around (\ref{eq:LargeSSD}) that the leading moment $n=1$ is simply given by the normalization of the corresponding basis state.

For such functions, one can prove that the order $(N,N)$ diagonal Pad\'e approximants $P_N^N(x)$ and the order $(N,N+1)$ off-diagonal Pad\'e approximants $P_{N+1}^N(x)$ are upper and lower bounds, respectively, on the exact function at $x>0$,  and moreover that they monotonically decrease and increase, respectively, with increasing $N$ \cite{bender2013advanced}.  In equations, $P_1^0(x) \le P_2^1(x) \le \dots \le f(x) \le \dots \le P_2^2(x) \le P_1^1(x)$, when $x>0$.   Therefore if for some $N$, $P_N^N(x)$ is very close to $P_{N+1}^N(x)$, then they are guaranteed to give a good approximation to the true function on the positive real axis.  Of course, in our case we have an additional source of uncertainty, which is that LCT does not give us the exact Taylor coefficients around $s=0$.  However, as long as we limit ourselves to Pad\'e approximants that use only the Taylor coefficients that have converged as a function of $\Delta_{\rm max}$ well enough to do an accurate extrapolation to $\Delta_{\rm max}=\infty$, then we will have an accurate calculation of these Pad\'e approximants, and in that case they will provide bounds on the true time-ordered correlators.  Finally, our ultimate goal for our bootstrap application is to obtain the time-ordered correlators at $s>4m^2$, i.e. $x<0$, and in this case we do not know of any results that say the true result is bounded by Pad\'e approximants.  However, one can still show \cite{bender2013advanced} that the diagonal and off-diagonal sequences $P_N^N(z)$ and $P_{N+1}^N(z)$ converge in the entire cut plane, and if their $N\rightarrow \infty$ limits are identical then they converge to the true time-ordered correlator near the branch cut as well.

\section{Form Factors}
\label{sec:LCT_FF}

In this section, we explain how to compute form factors in the LCT approach. In section \ref{sec:FF_LCT_subsection}  we introduce the formalism and show an explicit example. We proceed in section \ref{eq:PertAnalysis} by checking our results in perturbation theory. In section \ref{sec:FF_improvements} we show how one can significantly improve the precision of our numerical results by using various approximations of the raw data. We present our final numerical results for the form factor of the trace of the stress tensor  in the $\phi^4$ model (in the  $s<0$ kinematic regime) in section \ref{sec:FF_final_results}. We consider the analytic continuation of these results to the $s>0$ kinematic regime in section \ref{sec:FFPositives}.

\subsection{Computing Form Factors}
\label{sec:FF_LCT_subsection} 
 
Form factors are the matrix elements of local operators in a basis of asymptotic states.  In LCT, they are more difficult to compute than spectral densities because we do not have direct access to asymptotic states, only to eigenstates of the Hamiltonian.  However, there is one kind of asymptotic state that we can immediately calculate, namely the stable single-particle states $|m, \vec p_i\,\>$, where $\vec p_i$ is the spatial momentum of the $i$th particle. The single-particle state is an eigenstate of the Hamiltonian, and also both an `in' and `out' asymptotic state. The single-particle states obey the ``mass-shell'' condition  $-p_i^2=m^2$. As a result their energy reads $p^0 = \sqrt{m^2+\vec p^{\,2}}$.

In the two-dimensional lightcone coordinates, we can label the single-particle states by $p_{i-}$ and the other coordinate $p_{i+}$ is fixed by the ``mass-shell'' condition as
\begin{equation}
p_{i+} = \frac{m^2}{2p_{i-}}.
\end{equation}
Thus, we denote the single-particle states by $|m,p_{i-}\rangle$ in the lightcone coordinates.
In 2d ,it is often more convenient to work with the rapidity variable $\theta_i$ defined via
\begin{equation}
\label{eq:rapidities}
p_i^0 = m \cosh\theta_i,\qquad
p_i^1 = m \sinh\theta_i.
\end{equation}
In terms of the lightcone coordinates, we simply have
\begin{equation}
\theta_i=\log \left(\frac{m}{p_{i-}}\right).
\end{equation}

In LCT, by diagonalizing the Hamiltonian $P_+$ (or equivalently the mass-squared operator $M^2$), we obtain a set of eigenstates. Selecting the eigenstate with the lowest eigenvalue, we obtain the single-particle state $|m, p_-\rangle$. Using it we can compute the following two-particle form factor of a local operator $\cO$ with Lorentz spin $\ell$ in 2d
\be
F_{1,1}^{\CO}(\theta_1, \theta_2) = e^{- \frac{\ell}{2} (\theta_1+\theta_2)} \CF_{1,1}^{\CO}(\theta) \equiv \<m,p_{1-} | \CO(0) |m, p_{2-}\>,
\label{eq:InvariantFF}
\ee
where $\theta\equiv \theta_1-\theta_2$. The first equality holds due the covariance to under boosts. For convenience, let us discuss here various variables one can use to describe the two-particle form factors. Instead of $\theta$, one can use either the $s$ or $t$ variables defined by
\begin{equation}
\label{eq:st}
s\equiv -(p_1+p_2)^2,\qquad
t\equiv -(p_1-p_2)^2,\qquad
t=4m^2-s.
\end{equation}
Plugging \eqref{eq:rapidities} into \eqref{eq:st}, we find that
the $s$ and $t$ variables are related to the rapidity variable $\theta$ as
\begin{equation}
s= 4m^2 \cosh^2(\theta/2),\qquad
t= -4m^2 \sinh^2(\theta/2).
\end{equation}
Another variable that will be useful for us is
\begin{equation}
X \equiv \frac{p_{1-}}{p_{2-}}.
\end{equation}
Setting $p_-=p_{1-}+p_{2-}$, one can show that
\begin{equation}
\label{eq:s_X_rel}
s= \frac{m^2 p_-^2}{p_{1-}p_{2-}} = \frac{m^2 (1+X)^2}{X}.
\end{equation}

We will be particularly interested in the case where $\CO$ is the stress-tensor.  In lightcone quantization, the component $T_{--}$ is not only holomorphic (i.e. depends only on $x^-$, not $x^+$) in the CFT limit, but moreover it is simply $T_{--} = -(\partial_- \phi)^2$ even in the presence of the relevant deformations $\phi^2$ and $\phi^4$.  We can relate its form factors to that of the trace of the stress tensor
\begin{equation}
\Theta \equiv T_\mu^\mu = 2 T_{+-}.
\end{equation}
This is done by using the Ward identity
\be
[P_+, T_{--}] + [P_-, \frac{1}{2} \Theta] =0
\ee
which leads to the following simple relation
\begin{equation}
\< m, p_{1-}| \Theta(0) | m, p_{2-} \> = e^{ \theta_1 + \theta_2} \< m, p_{1-} | T_{--}(0) | m, p_{2-} \>.
\end{equation}
By inspection of (\ref{eq:InvariantFF}) and the fact that $T_{--}$ and $\Theta$ transform under boosts like $\ell=2$ and $\ell=0$ respectively, we see that
\be
\CF_{1,1}^{\Theta}(\theta) = \CF_{1,1}^{T_{--}}(\theta),
\ee
so when we compute the form factor, we may use $T_{--}$ or $\Theta$ depending on which is easier in context.

Let us discuss in more detail how we compute $\CF_{1,1}^{\CO}$ in LCT.
Diagonalizing the Hamiltonian gives us the eigenstates in the form \eqref{eq:EstatesInLCTBasis}. The eigenstate corresponding to the smallest $\mu_i^2$ gives us the one-particle state $|m,p_-\>$. For clarity we denote the coefficients $C_{\CO_j}^{\mu_i^2}$ in  \eqref{eq:EstatesInLCTBasis} by $c_j$ in the case of one-particle states. Thus, we can write
\be
|m,p_- \> = \sum_j c_j | \CO_j, p\>.
\ee
Inserting this expression in the definition of $\CF_{1,1}^\CO$ given by \eqref{eq:InvariantFF}, we find
\begin{align}
	\CF_{1,1}^{\CO}(\theta_1,\theta_2) &= \sum_{j, j'} c^*_j c_{j'} \< \CO_j, p_1 | \CO(0) | \CO_{j'}, p_2 \>,
\end{align}
where due to \eqref{eq:states} the matrix element reads as
\begin{align}
\< \CO_j, p_1 | \CO(0) | \CO_{j'}, p_2 \>
& = \sum_{j, j'} \frac{1}{N_{\CO_{j}} N_{\CO_{j'}}}  \int d^2 x d^2 x' e^{i (p_1 \cdot x - p_2 \cdot x' )} \< \CO_{j}(x) \CO(0) \CO_{j'}(x')\>.
\end{align}
The three-point function on the RHS above is a Wightman three-point function of primary operators in a CFT, and therefore is fixed by conformal invariance up to an overall constant OPE coefficient.  The integrals over $x$ can be computed in closed form \cite{Anand:2019lkt}. In this work, we will  only need the result in the specific case where $\CO, \CO_j$ and $\CO_{j'}$ are all holomorphic. We thus have
\begin{align}
\nn
\< \CO_j, p_1 | \CO(0) | \CO_{j'}, p_2 \> &=
C_{\CO j j'} \cdot (p_{1-} p_{2-})^{\frac{h_{\CO}}{2}} X^{h_j - \frac{h_\CO}{2}}\\ &\times 4\pi \frac{\sqrt{\Gamma(2h_j)\Gamma(2h_{j'})}}{\Gamma(h_j+h_{j'} +h_\CO-1)}  P_{h_\CO+h_{j'} - h_j -1}^{(2h_j-1, 1-2h_\CO)}(1-2X) ,
\label{eq:3pt}
\end{align}
where $0\le p_{1-} \le p_{2-}$. The corresponding formula for $0 \le p_{2-} \le p_{1-}$ follows easily by taking the Hermitian conjugate to swap the bra and ket state.
Here $C_{\CO j j'}$ is the $\CO \CO_j \CO_{j'}$ OPE coefficient in the UV CFT (without the relevant deformation), $P_n^{(\alpha, \beta)}$ are the Jacobi polynomials.
Putting it all together, we obtain a formula for the form factor $\CF_{1,1}^\CO$ when $\CO$ and the basis operators $\CO_j$ are holomorphic:
\begin{equation}
\boxed{\CF_{1,1}^\CO(\theta) = \sum_{j,j'} c_j^* c_{j'} C_{\CO j j'} X^{h_j - \frac{h_\CO}{2}} \frac{4\pi m^{h_\CO} \sqrt{\Gamma(2h_j)\Gamma(2h_{j'})}}{\Gamma(h_j+h_{j'} +h_\CO-1)}  P_{h_\CO+h_{j'} - h_j -1}^{(2h_j-1, 1-2h_\CO)}(1-2X) .
\label{eq:TruncFormFactorBasic}}
\end{equation}
We explain in Appendix \ref{app:OPE} how we compute the OPE coefficients efficiently in the 2d free massless scalar theory, using a generalization of the methods from \cite{Anand:2020gnn}.

In principle, if we know $\CF_{1,1}^{\CO}(\theta)$ exactly, then we can use crossing symmetry and analyticity to also obtain the following {\it two} particle form factor:
\be
\CF_{2,0}^{\CO}(\theta) \equiv e^{\frac{\ell}{2} (\theta_1+\theta_2)} {}_{\rm out}\<m,p_{1-}; m,p_{2-} | \CO(0)|{\rm vac}\> =\CF_{1,1}^\CO(\theta+i \pi).
\ee
However, in practice we will only know $\CF_{1,1}^{\CO}$ at real values of $\theta$, and we will not know it exactly.  In particular, we cannot simply perform the analytic continuation of $\theta$ term-by-term in the sum in (\ref{eq:TruncFormFactorBasic}). One way to see why analytically continuing equation (\ref{eq:TruncFormFactorBasic}) in $\theta$  to obtain the $\CF_{2,0}^\CO$ form factor cannot be as simple as analytically continuing each individual term in the sum is that, for even $h_{\CO}$, each term is manifestly a polynomial in $X$, and this ratio goes from being real and positive to being real and negative (as can be seen from \eqref{eq:s_X_rel}) when $\theta \rightarrow \theta + i \pi$ to turn the out state into an in state.  Therefore, the sum (\ref{eq:TruncFormFactorBasic}) is a sum over real numbers, whereas the $\CF_{2,0}^\CO$ form factor, with two in-states, will in general have a complex phase related to physical scattering processes.

\subsubsection{Illustrative Example} 
As an illustration, let us consider the case where we take a very small $\Delta_{\rm max}=5$.  There are only four $\mathbb{Z}_2$-odd primaries in the UV CFT with $\Delta \le \Delta_{\rm max}$:
\be
\CO_1 \propto \partial \phi, \quad \CO_2 \propto (\partial \phi)^3, \quad \CO_3 \propto (6 \partial^3 \phi (\partial \phi)^2 - 9 (\partial^2 \phi)^2 \partial \phi),\quad
\CO_4  \propto  (\partial \phi)^5.
\ee
 The Hamiltonian matrix elements from the mass term $\CL \supset - \frac{m_0^2}{2} \phi^2$ and interaction term $\CL \supset - \frac{\lambda}{4!} \phi^4$ are $M^2 = -P^2=2P_+ P_- = m_0^2 M^2_{\phi^2} + \lambda M^2_{\phi^4}$ with
 \be
 M^2_{\phi^2} =  \left(
\begin{array}{cccc}
 1 & 0 & 0 & 0 \\
 0 & 15 & 4 \sqrt{3} & 0 \\
 0 & 4 \sqrt{3} & 27 & 0 \\
 0 & 0 & 0 & 45 
\end{array}
\right), 
\qquad 
M^2_{\phi^4} = \left(
\begin{array}{cccc}
 0 & \frac{\sqrt{5}}{4 \pi } & \frac{\sqrt{15}}{8 \pi } & 0 \\
 \frac{\sqrt{5}}{4 \pi } & \frac{15}{4 \pi } & \frac{\sqrt{3}}{\pi } & \frac{\sqrt{105}}{2\pi}\\
 \frac{\sqrt{15}}{8 \pi } & \frac{\sqrt{3}}{\pi } & \frac{33}{8 \pi } & 0 \\
 0 & \frac{\sqrt{105}}{2\pi} & 0 & \frac{45}{2\pi}
\end{array}
\right)
.
\ee

We can now diagonalize $M^2$ numerically for any given value of $\bar\lambda$. As a result of this diagonalisation at, for example,  $\bar\lambda =1 $ we obtain the following eigenstate
\begin{equation}
\label{eq:eigenstate_1PS}
|p \> = 0.99994 | \CO_1, p\> -0.01036 |\CO_2,p\> - 0.00280 | \CO_3, p\> + 0.00033 | \CO_4,p\>,
\end{equation}
which corresponds to the lowest eigenvalue $m^2\approx 0.99773$ (physical mass). The state \eqref{eq:eigenstate_1PS} is the one-particle states with mass $m$.
When $\bar\lambda$ is small one can perform the above diagonalization for a generic value of $\bar\lambda$ order by order in $\bar\lambda$. For instance one has\footnote{\label{foot:time-indep_vs_diag}Alternatively this expression can also be derived using time-independent perturbation theory, see appendix \ref{app:time-independent_perturbation_theory}. For larger size of the mass-squared matrix $M^2$, the latter becomes much more efficient for computing the perturbative corrections. We will employ this technique in the next section.}
\be\label{eq:PertOnePart}
|p\> = | \CO_1,p \> - \bar{\lambda} \left( \frac{5 \sqrt{5}}{316 \pi } | \CO_2,p \> +\frac{3 \sqrt{15}}{1264 \pi } | \CO_3,p\> \right) + \CO(\bar\lambda^2).
\ee
Notice that the state $| \CO_4,p\>$ does not enter the expression \eqref{eq:PertOnePart} at linear order in $\bar\lambda$. For $\bar\lambda=1$, \eqref{eq:PertOnePart} matches well the numerical expression \eqref{eq:eigenstate_1PS}.
 
 The last missing ingredient for obtaining the form factor of trace of the stress tensor is the OPE coefficients of the stress tensor with other operators.  These can be computed by hand or using the methods in appendix  \ref{app:OPE}.  We find\footnote{Following a common abuse of notation, we have denoted $R\CO(0)R$ by ``$\CO(\infty)$'', where $R$ is a conformal inversion. The operators $\CO_j$ are normalized by the Zamolodchikov metric, i.e. $\< \CO_j(\infty)   \CO_{j}(0) \> \equiv 1$.  By contrast, the stress tensor is defined  here as $T_{--} \equiv  - (\partial_- \phi)^2$, where $\< \phi(x) \phi(0)\> \equiv -\frac{1}{4\pi} \log x$.}
\be
C_{Tj j'} = \< \CO_j(\infty) T_{--}(1)  \CO_{j'}(0) \> = -(4\pi)^{-1}
\left(
\begin{array}{cccc}
 2 & -\sqrt{6} & -2 \sqrt{\frac{6}{7}} & 0 \\
 -\sqrt{6} & 6 & \frac{12}{\sqrt{7}} & -2\sqrt{5}\\
 -2 \sqrt{\frac{6}{7}} & \frac{12}{\sqrt{7}} & 10 & 0 \\
 0 & -2\sqrt{5} & 0 & 10
\end{array}
\right) .
\ee

Plugging all the above results into equation \eqref{eq:TruncFormFactorBasic}, we obtain the final expression for the form factor of the trace of the stress tensor. For example, for $\bar\lambda=1$, using the state in (\ref{eq:eigenstate_1PS}),  we get
\begin{multline}\label{eq:FF_numeric_LCT}
m^{-2}\CF_{1,1}^\Theta(X) +2 = 0.000804 (1-X)^2 - 0.0118 P_3^{(1,-3)}(1-2X)\\
 +0.000078 P_4^{(1,-3)}(1-2X) - 0.00437 P_5^{(1,-3)}(1-2X).
\end{multline}
Notice that $P_n^{(1,-3)}(1-2X)$ vanish like $\sim (1-X)^3$ at $X=1$ for $n\ge 3$. At small coupling instead, we can use the perturbative one-particle state (\ref{eq:PertOnePart}), and we get analytically
\begin{align}
\label{eq:FF_perturbative_LCT}
\frac{m^{-2}\CF_{1,1}^{\Theta}(X)+2}{\bar\lambda/4\pi} &= -\frac{1}{158} \left( 25 P_3^{(1,-3)}(1-2X) + 9 P_5^{(1,-3)}(1-2X) \right) +\CO(\bar\lambda).
\end{align}

\begin{figure}[t!]
\begin{center}
\includegraphics[width=0.48\textwidth]{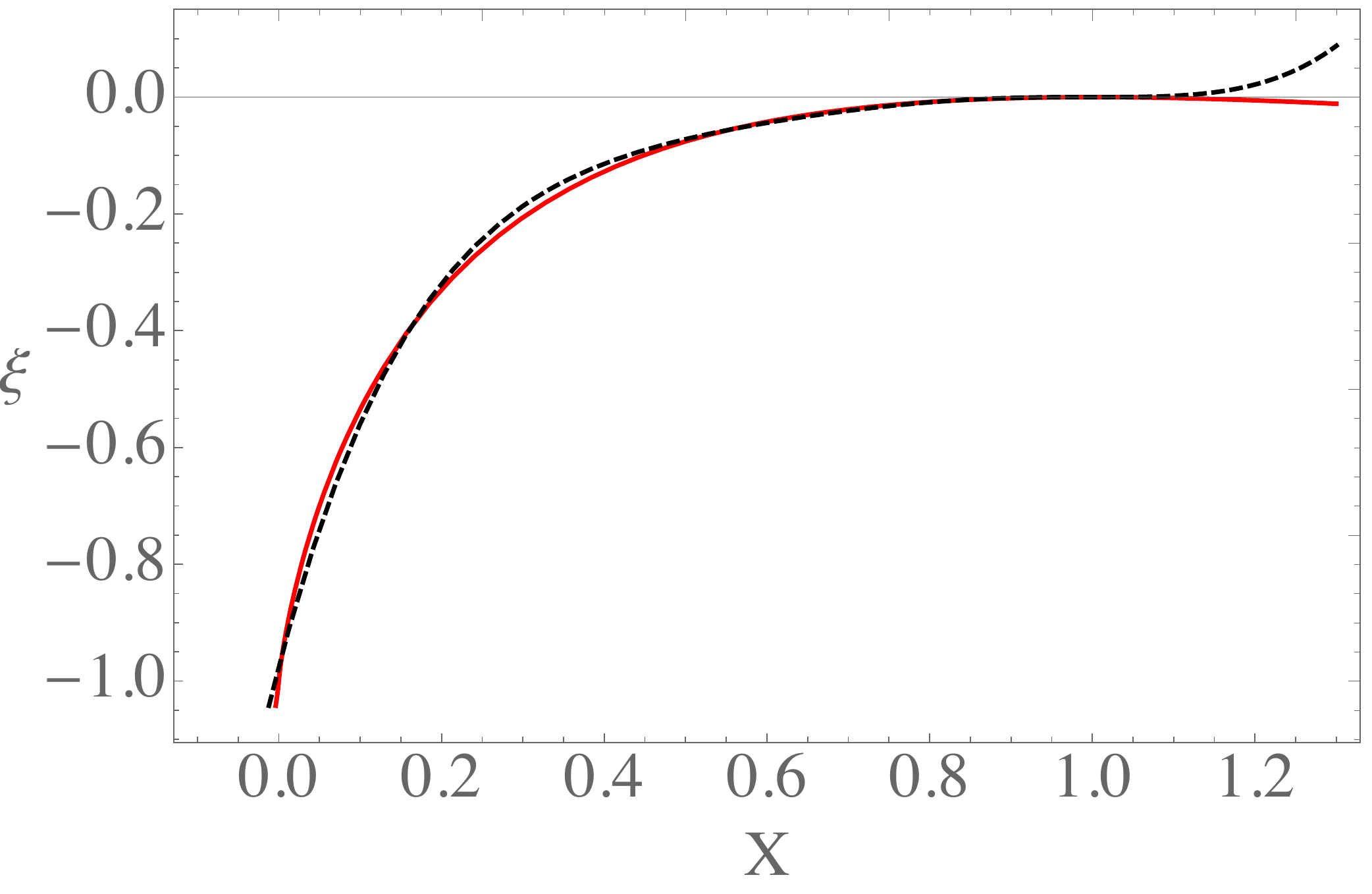}
\caption{Comparison of $\xi(X)\equiv\frac{m^{-2}{\cal F}_{1,1}^\Theta(X) +2}{\bar\lambda/4\pi}$ computed using the LCT approach with $\Delta_{\rm max}=5$ ({\it black, dashed}) and using perturbation theory ({\it red, solid}) at leading order in $\bar\lambda$.
}.
\label{fig:warmupFF}
\end{center}
\end{figure}

The perturbative analytic expression for the form factor of the trace of the stress tensor can be computed from Feynman diagrams. The result up to two-loop (order $\bar \lambda^2$) is given in \eqref{eq:FF_pert}. Let us write it here again for convenience. At linear order in $\bar\lambda$, it reads
\be
\label{eq:FF_perturbative}
\frac{m^{-2}\CF_{1,1}^{\Theta}(X)+2}{\bar\lambda/4\pi} = \frac{2 X \log (X)}{X^2-1}-1 + \CO(\bar\lambda).
\ee
We compare the LCT result \eqref{eq:FF_perturbative_LCT} with the analytic result  \eqref{eq:FF_perturbative} in Fig. \ref{fig:warmupFF}. We find an excellent agreement.
 
\subsection{Analysis of Residuals}
\label{eq:PertAnalysis}

In this section we analyze in more detail the residual errors in our computation of the form factor. 

\begin{figure}[t]
\begin{center}
\includegraphics[width=0.44\textwidth]{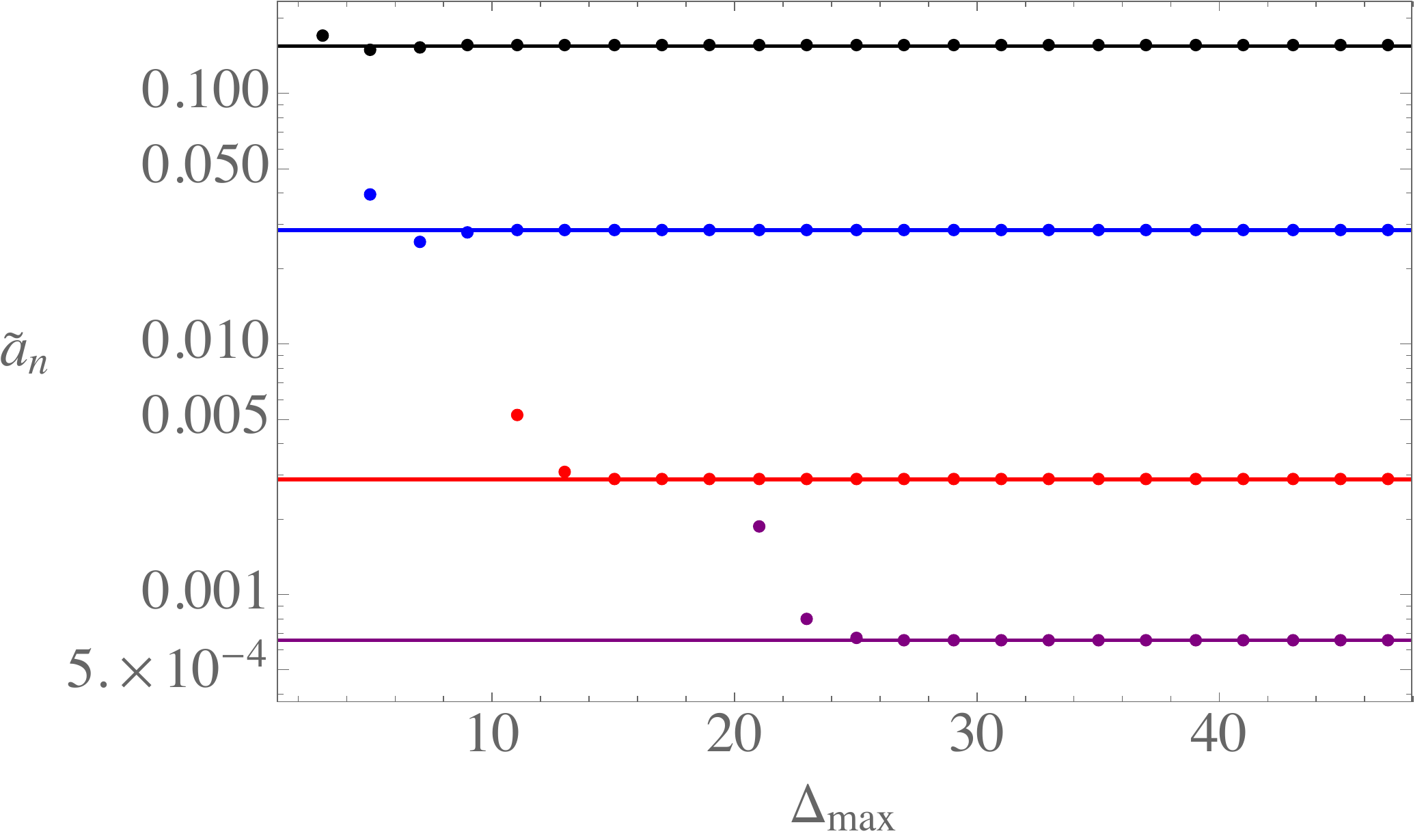}
\includegraphics[width=0.48\textwidth]{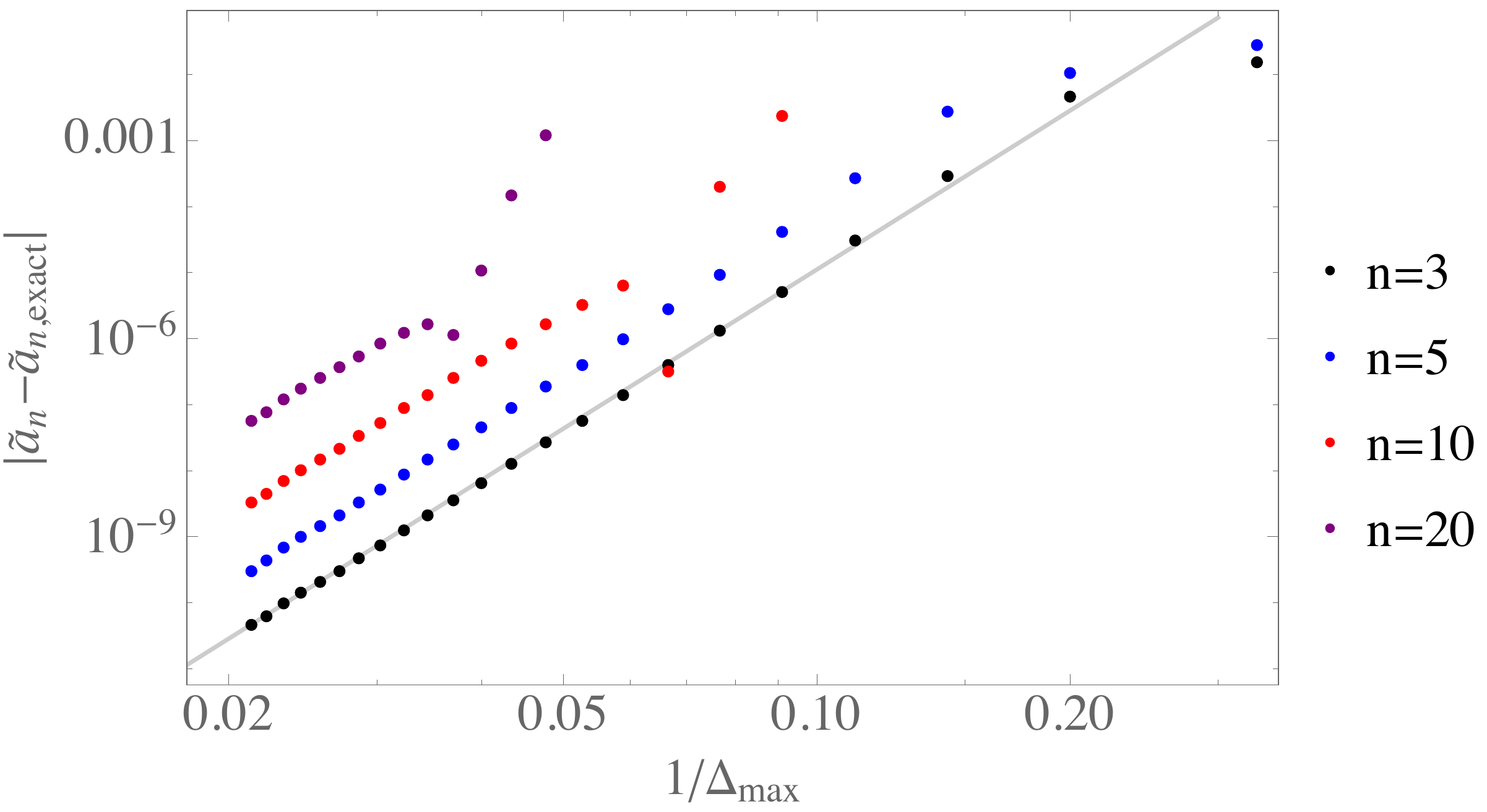}
\caption{Convergence of the $\tilde{a}_n$ coefficients of the normalized Jacobi polynomials $\tilde{P}_n^{(1,-3)}$ for the one-loop form factor ($\xi(X)=\sum_n \tilde{a}_n \tilde{P}_n^{(1,-3)}$).  {\it Left:} Plot of coefficients as a function of $\Delta_{\rm max}$.  The straight lines are the exact values computed from (\ref{eq:1loopJacobiExpansion}), shown for comparison.  {\it Right:} Absolute value of the difference between truncation result for $\tilde{a}_n$ and the exact result, as a function of $1/\Delta_{\rm max}$.  The line $(2.4/\Delta_{\rm max})^8$ is an approximate numeric fit to the case $n=3$, shown for comparison. }
\label{fig:JacobiConvergence}
\end{center}
\end{figure}

Let us start from the  ``perturbative'' LCT expression \eqref{eq:FF_perturbative_LCT}. It tries to build  \eqref{eq:FF_perturbative} as a sum over Jacobi polynomials. To see this precisely, expanding \eqref{eq:FF_perturbative} in a series of Jacobi polynomials, one has
\begin{equation}\label{eq:1loopJacobiExpansion}
\frac{2 X \log (X)}{X^2-1}-1 = \sum_{n=3}^\infty a_n P_n^{(1,-3)}(1-2X),
\end{equation}
where the values of the first three coefficient read as\footnote{These coefficients can be computed straightforwardly since the polynomials  $P_n^{(1,-3)}(1-2x)$ are orthogonal with respect to the inner product $(f,g)=\int_0^1 dx \frac{x}{(1-x)^3} f(x) g(x)$ and have norm  $\sqrt{(P_n^{(1,-3)} , P_n^{(1,-3)})} =  \sqrt{\frac{n+1}{(n-2) (2 n-1)}}$.}
\begin{equation}
a_3 = \frac{5(-10+\pi^2)}{4} \approx -0.162994,\quad
a_4 = 0, \quad
a_5 = -533 + 54 \pi^2 \approx -0.04136.
\end{equation}
These are fairly close to the ones of the truncation result \eqref{eq:FF_perturbative_LCT}, with which are
\begin{equation}
a_3=-25/158\approx-0.158,\quad
a_4=0,\quad
a_5=-9/185\approx-0.057.
\end{equation}
These were obtained at $\Delta_{\rm max}=5$. When we increase $\Delta_{\rm max}$\footnote{In the perturbative regime at larger values of $\Delta_{\rm max}$ in this section, we use time-independent perturbation theory discussed in appendix \ref{app:time-independent_perturbation_theory} in order to obtain the eigenstates instead of diagonalizing the mass-square matrix $M^2$. See also footnote \ref{foot:time-indep_vs_diag}.}, these coefficients approach the exact values fairly quickly -- numerically, for instance, the error on $a_3$ behaves approximately like
\begin{equation}
|a_3-a_{3,\rm exact}| \approx \left(\frac{2.4}{\Delta_{\rm max}}\right)^{8}.
\end{equation}
In Fig. \ref{fig:JacobiConvergence}, we show the convergence for $n=3,5,10,20$ for the coefficients $\widetilde{a}_{n}$ of the normalized Jacobi polynomials $\tilde{P}_n^{(1,-3)}$ defined as
\begin{equation}
\widetilde{a}_{n} \equiv \frac{a_{n}}{\sqrt{\frac{(n-2)(2 n-1)}{n+1}}},\qquad
\tilde{P}_n^{(1,-3)}\equiv \sqrt{\frac{(n-2) (2 n-1)}{n+1}} P_n^{(1,-3)}.
\end{equation}
However, note that for larger values of $n$, one must reach higher values of $\Delta_{\rm max}$ before the asymptotic convergence rate sets in.  This is evident already from the fact that the Jacobi polynomials of high order do not even begin to appear in the LCT formula until large values of $\Delta_{\rm max}$.  

Summarizing, in practice, at any finite $\Delta_{\rm max}$, LCT will  give us an accurate estimate for some finite number of coefficients $a_n$ and a poor estimate for the remaining infinite set of coefficients in the decomposition of the form factor as a sum over Jacobi polynomials.  The sum over Jacobi polynomials is  absolutely convergent to the true form factor on the interval $0<X<1$.  However, the same cannot be said of the derivative of the form factor (with respect to, say, $X$).  In general, as we increase $\Delta_{\rm max}$, the  resulting form factor approaches the true form factor with a residual whose amplitude is decreasing but with increasingly oscillatory behavior.  If one wants to analytically continue our result beyond the real line segment $0< X< 1$, one needs to find some way to remove these oscillatory residuals.  

As the values of the coefficients $a_n$ for small $n$ stabilize, most of the error in the form factor will come from the error in the $a_n$ coefficients with large $n$.  To get a more concrete sense of the shape of their contribution, we can look at the asymptotic formula for $P_n^{(1,-3)}$ at large $n$:
\be
\frac{X^{\frac{3}{4}}}{(1-X)^{\frac{5}{4}}}
 P_n^{(1,-3)}(1-2X) = -\frac{\cos \left[\pi  \Big(  ( n-\frac{1}{2})\varphi+\frac{1}{4}\Big)\right]}{\sqrt{\pi n }} 
 + \CO(n^{-3/2}),
\ee
where $X= \sin^2 (\pi \varphi)$. 
  The largest source of error will typically come from the most ``recently'' added Jacobi polynomials, i.e. those with large $n$ that appear in the truncation result but whose coefficients have not converged as well as those with smaller values of $n$.  So most of the error will be highly oscillatory, with a period of roughly $\sim \frac{1}{\Delta_{\rm max}}$ in the variable $\varphi$.  By contrast, the ``slowly-varying'' part of the form factor will be much more accurate.  Moreover, the slowly varying part of the form factor is the part that behaves well under analytic continuation.

\begin{figure}   
 \centering
 \includegraphics[height=4.5cm]{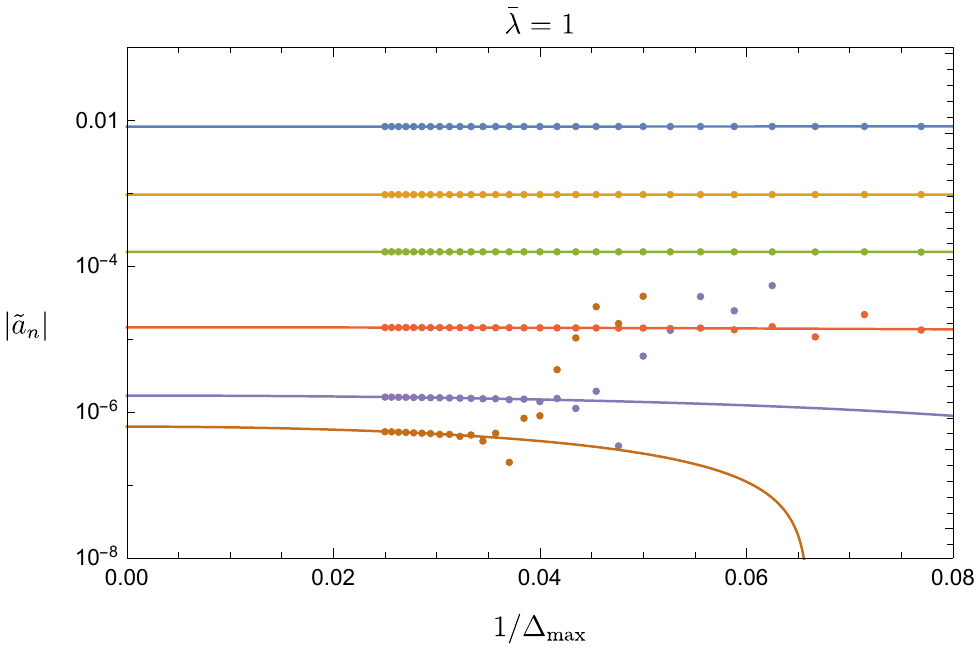} 
        \includegraphics[height=4.5cm]{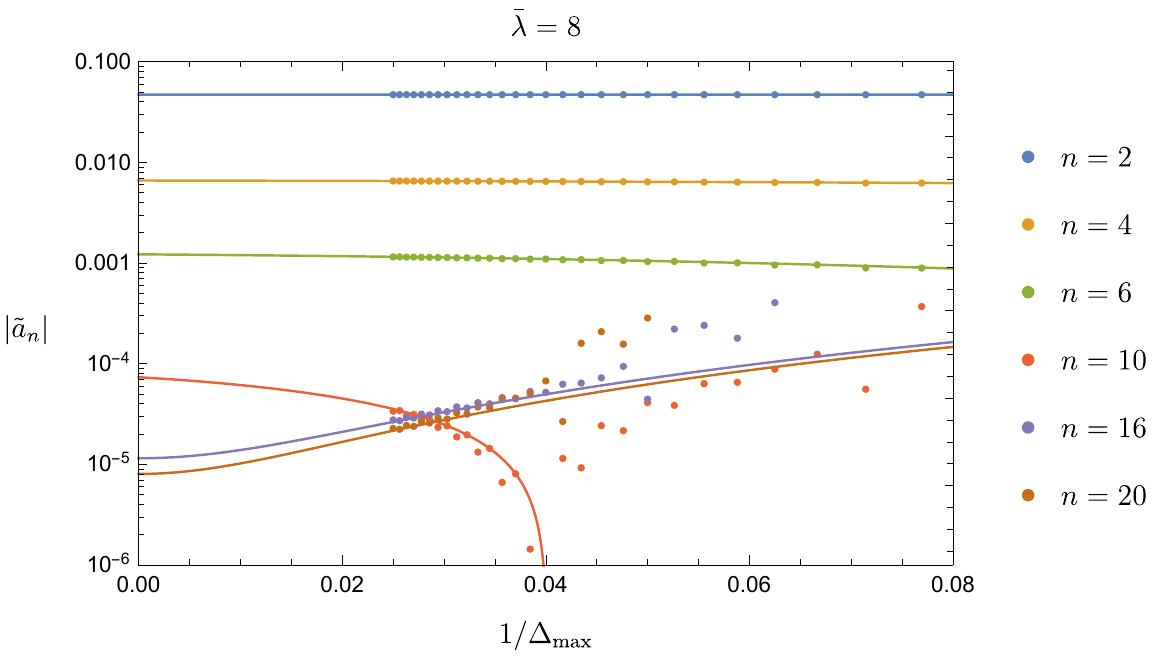} 
        
 \includegraphics[height=4.5cm]{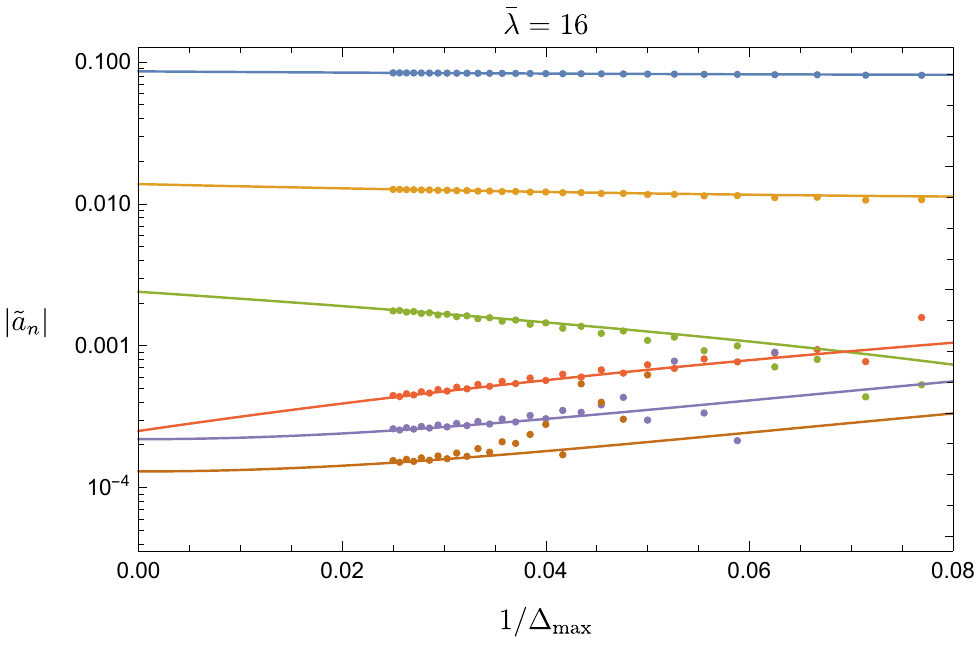} 
        \includegraphics[height=4.5cm]{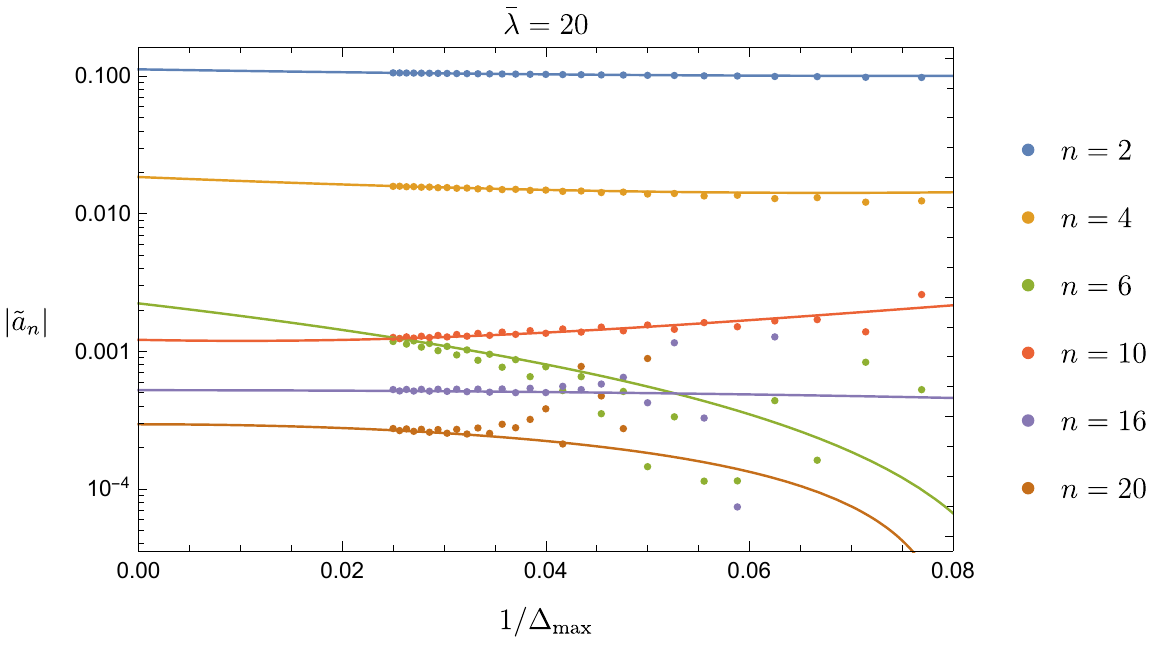} 
 \caption{Convergence of the $\tilde{a}_n$ coefficients of normalized Jacobi polynomials $\tilde{P}_n^{(1,-2)}$ for the non-perturbative form factor for various $\bar\lambda$. Here we show the coefficients as a function of $x=1/\Delta_{\rm max}$.  We extrapolate these coefficients to infinite $\Delta_\text{max}$ by using a quadratic function $a +b x +c x^2$ to fit these coefficients (actually, for $n>16$, we used $a +c x^2$, which seems to work better). The solid lines are the results of the fits. The convergence of these coefficients becomes worse for larger $\bar\lambda$, and in this paper, we have gone up to $\bar\lambda=20$, and for most cases, we take the sum of $\tilde{a}_n\tilde{P}_n^{(1,-2)}$ up to $n=18$ as an approximation to the form factors. 
 }
 \label{fig:ConvCoeffLambda1}
\end{figure}

We conclude this section by pointing out that the numerical non-perturbative LCT result \eqref{eq:FF_numeric_LCT} for $\bar\lambda=1$ is numerically very close to the ``perturbative'' LCT result \eqref{eq:FF_perturbative_LCT}, since $\bar\lambda=1\ll 4\pi$ is in the perturbative regime.   The main qualitative difference is that now the form factor approaches $-2m^2$ like $(1-X)^2$ as $X \sim 1$. This feature makes the Jacobi polynomials $P_n^{(1,-2)}(1-2X)$ a more appropriate basis than $P_n^{(1,-3)}(1-2X)$, since they each individually have this behavior for $n\ge 2$. Therefore, to get a more accurate result for the form factors, we convert the different Jacobi polynomials in equation (\ref{eq:TruncFormFactorBasic}) to $P_n^{(1,-2)}(1-2X)$ using some Jacobi polynomial identities. See appendix \ref{app:jacobi} for more details. As a result, our final form factor is written as
\begin{equation}
\label{eq:final_raw_from_FF}
m^{-2}\CF_{1,1}^\Theta(X)+2 =
\sum_{n} \widetilde{a}_{n} \widetilde{P}_{n}^{(1,-2)}(1-2 X),
\end{equation}
where $\tilde{P}_n^{(1,-2)}$s are the  normalized Jacobi polynomials defined as
\begin{equation}
\tilde{P}_n^{(1,-2)} = \left( \frac{2n(n-1)}{n+1} \right)^{1/2} P_n^{(1,-2)}(1-2X).
\end{equation}
In Fig. \ref{fig:ConvCoeffLambda1}, we show the convergence of the coefficients at various $\bar\lambda$. For each $\bar\lambda$, we computed the coefficients $\tilde a_n$ for each $\Delta_\text{max}$ up to $\Delta_\text{max}=40$, and extrapolated them to $\Delta_\text{max}=\infty$ as a function of $1/\Delta_\text{max}$\footnote{In fact, as in the case for extrapolating the coefficients in the time-ordered two-point function in section (\ref{sec:SD_Pade}), we found that the convergence is better if we fix the mass gap to be the same for each value of $\Delta_{\rm max}$, which requires dialing the coupling $\bar\lambda$ slightly as a function of $\Delta_{\rm max}$ to keep the gap fixed.}. 

In general, for fixed $n$, $\tilde a_n$ becomes larger as we increase $\bar\lambda$. Therefore, to obtain more accurate result for the form factor with larger $\bar\lambda$, we will have to sum up to larger $n$. Unfortunately, for fixed $\bar\lambda$, the convergence becomes slower as we increase $n$. In this paper, we have settled on computing the sum in equation (\ref{eq:final_raw_from_FF}) up to $n=18$ for various $\bar\lambda$ values, which means that the result is less accurate for larger $\bar\lambda$. From figure \ref{fig:ConvCoeffLambda1}, one can see that assuming the extrapolation to $\Delta_\text{max}=\infty$ is accurate, discarding $\tilde a_n$ with $n\ge 19$ will introduce an uncertainty of order $\CO(10^{-6}))$ for $\bar \lambda=1$ and order $\CO(10^{-3})$ for $\bar \lambda=20$, since the Jacobi polynomial is $\widetilde{P}_{n}^{(1,-2)}(1-2 X)$ is order 1 in $0<X<1$ (it is slightly larger than order 1 near $X=0$).

\subsection{Improved Form Factor from Rational Approximations}
\label{sec:FF_improvements}

In this section we discuss two methods which allow one to improve the ``raw'' LCT results for the form factor. In practice we use both of them together when computing the final results. We will begin with a simple approach where we fit a rational function to our raw form factor in the regime $s<0$ where it is fairly accurate.  As we explain in more detail, the resulting rational fit will turn out to be more accurate than the original raw form factor.  However,  in the subsequent analysis in section \ref{sec:rational_approximation_analyticity}, we will show that with more work we can do significantly better if we perform such a rational fit {\it after} first improving the raw form factor in several ways that reduce the error in the $s<0$ regime.  These improvements take advantage of the structure of the truncation calculation itself, and in particular rely on the fact that truncation builds up the form factor as a sum over orthogonal polynomials.  In our final result, we perform the following sequence of steps: i) extract the coefficients of these orthogonal polynomials as a function of $\Delta_{\rm max}$, ii) for the coefficients that have started to converge, extrapolate them as a function of $\Delta_{\rm max}$ to $\Delta_{\rm max}=\infty$, otherwise we discard them if they have not started to converge, iii) extend the domain of convergence from the interval $0<X<1$ to an open subset of the complex plane by separating the raw form factor into the sum of two separate pieces, each which individually has a larger domain of convergence than their sum does, and iv) finally, fit a rational polynomial to the improved result of steps (i)-(iii).  We find that the final result is drastically improved at $s<0$, and works reasonably well even at $s>4m^2$ when we analytically continue.

\subsubsection{Rational Approximation from Direct Fit}
\label{sec:FFRationalFit}

In the previous subsection, we saw that the raw computation of the form factor $\CF_{1,1}^\CO$ in LCT produces a function that is quite accurate when the momenta of both the `in' and `out' particle are positive, but rapidly becomes completely incorrect as one analytically continues away from this region. Relatedly, the raw form factor is mainly contaminated by errors that are rapidly oscillation functions of the momentum ratio  $X$.  In this subsection, we will discuss a simple procedure for reducing these errors. This method will be particularly robust in the regime where the raw result is already accurate, and we are simply reducing small errors even further.  Outside of this regime, we will effectively be doing a kind of extrapolation which requires a bit of caution.  In a companion paper \cite{truncboot}, we will discard this extrapolation and replace it with bounds from the S-matrix/form factor bootstrap, where constraints from unitarity make the extrapolation more systematic.

The basic idea is to approximate the form factor as a rational function of a ``nice'' variable.  We find the approximate rational function by doing a fit over the range where the raw form factor is already a good approximation.  The ``nice'' variable we use here was already exploited in \eqref{eq:def_rho_1}. It is defined as
\begin{equation}
\label{eq:rho-variable}
\newRho \equiv
\lim_{\epsilon\rightarrow 0^+} \frac{2m-\sqrt{4m^2-s-i\epsilon}}{2m+\sqrt{4m^2-s-i\epsilon}}.
\end{equation}
Notice that \eqref{eq:def_rho_1} is simply the inverse of \eqref{eq:rho-variable}.
The ``rational approximation Ansatz'' can be written 
\be\label{eq:RationalApproxAnsatz}
\CF_{2,0}^\Theta(s) = -2m^2\left( 1+ \frac{\sum_{n=1}^{N} b_n \newRho^n}{1+\sum_{m=1}^N c_n \newRho^m}\right).
\ee
While the form of this Ansatz is identical to that of a Pad\'e approximation, we emphasize that its coefficients will be fixed by performing a fit rather than by matching Taylor series coefficients.
Clearly, the Ansatz depends on the choice of the number $2N$ of free parameters.  If $N$ is too small, the Ansatz becomes inaccurate because there are not enough terms to accurately reproduce the shape of the form factor, whereas if $N$ is too large then the best fit will simply reproduce the raw form factor and all the issues associated with it.  For some intermediate regime, however, the Ansatz is able to accurately reproduce the form factor while essentially ``smoothing out'' the high frequency oscillations that are causing most of the error, leading to a significant improvement.  In practice, we have found that $N=2$ is almost always too small, but $N=3$ already usually works well at reducing the errors.

\begin{figure}[t]
\begin{center}
\includegraphics[width=0.48\textwidth]{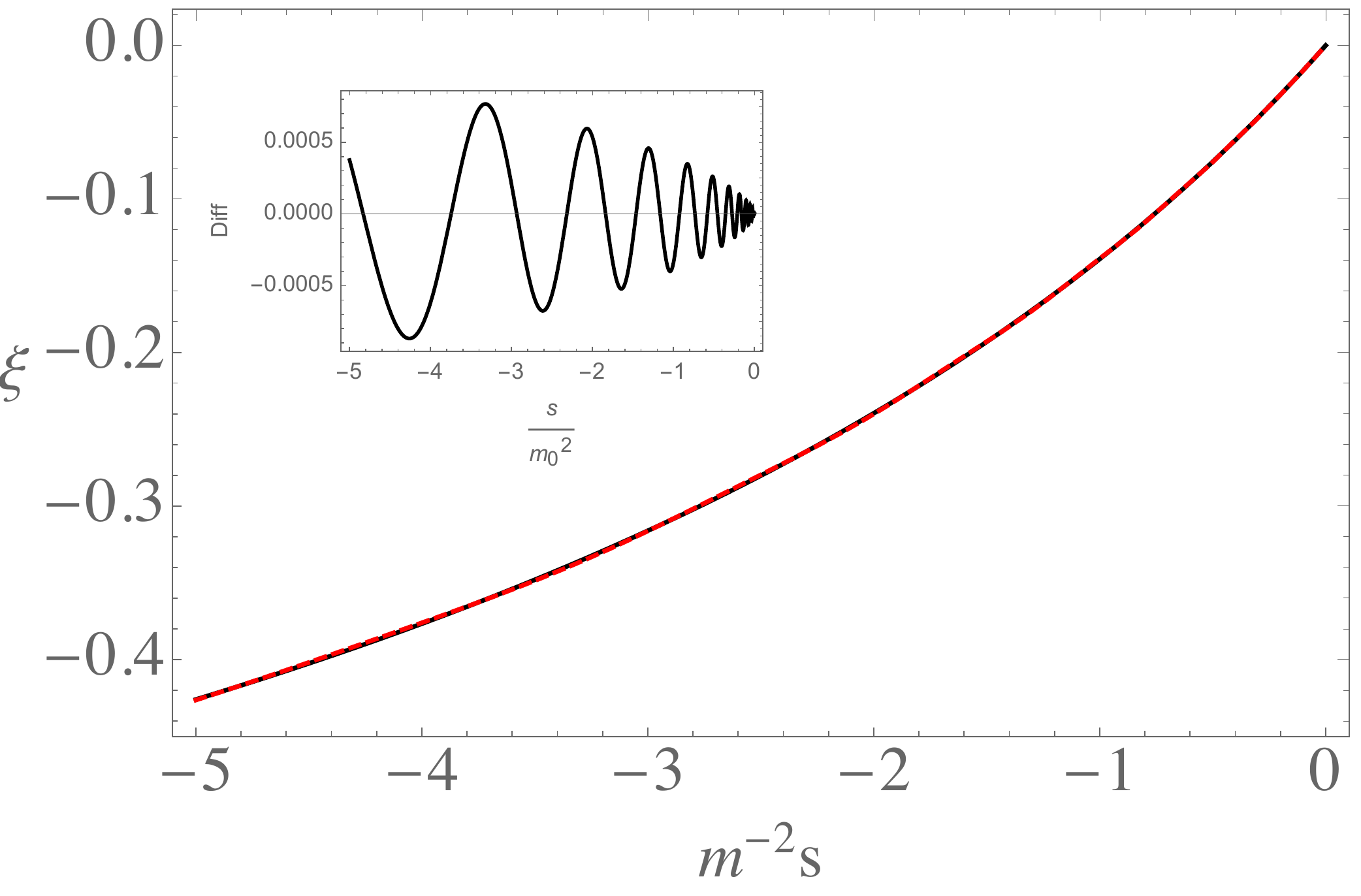}
\includegraphics[width=0.48\textwidth]{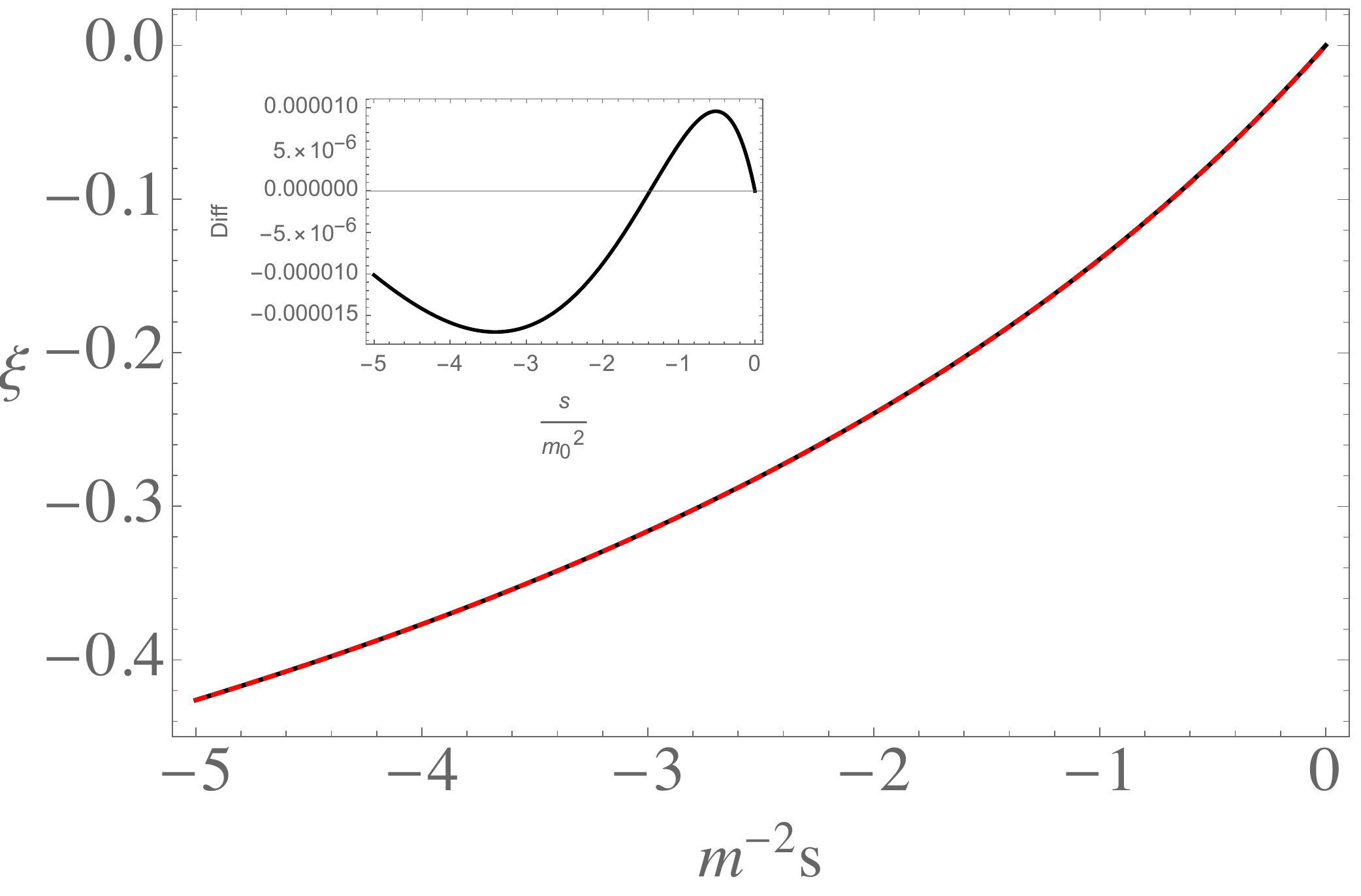}
\caption{Exact one loop form factor (black, thick) at $s<0$ compared with the LCT form factor  (red, dashed), and the residual errors (inset).  {\it Left}: ``Raw'' LCT result from direct computation expanded to $\CO(\bar\lambda)$.  {\it Right}: LCT result from performing a fit to  the ``raw'' result with a rational function (\ref{eq:RationalApproxAnsatz}) at $N=3$.  As can be seen from the insets, the rational function approximation removes the large ``high-frequency'' errors and reduces the overall error significantly. 
}
\label{fig:RationalApprox1}
\end{center}
\end{figure}

\begin{figure}[th!]
\begin{center}
\includegraphics[width=1.0\textwidth]{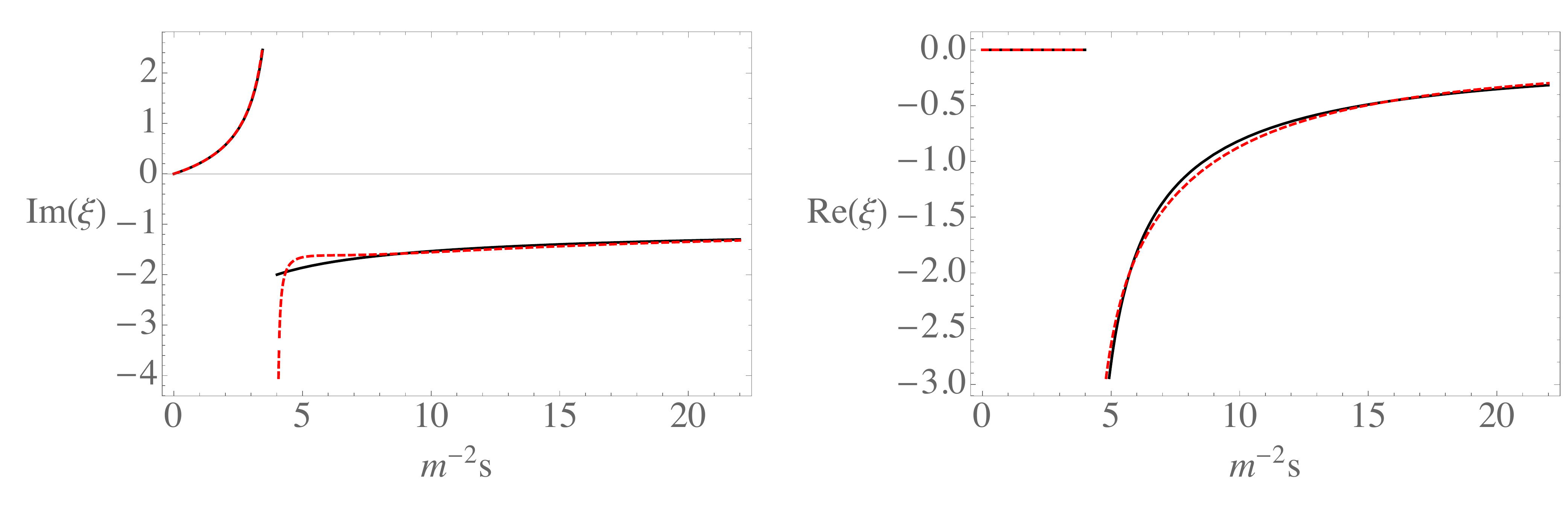}
\caption{Imaginary ({\it left}) and real ({\it right}) parts of the one-loop form factor, comparing the exact result (black, thick) to the result from performing a fit to the $\Delta_{\rm max}=40$ result with a rational function (\ref{eq:RationalApproxAnsatz}) at $N=3$ (red, dashed). 
}
\label{fig:RationalApprox2}
\end{center} 
\end{figure}

Let us demonstrate this method in the regime of small values of $\bar\lambda$. In this regime, we can simply compare the exact one-loop form factor \eqref{eq:FF_perturbative} with the LCT one-loop result computed using time-independent perturbative theory (see appendix \ref{app:time-independent_perturbation_theory} for more details). That is, we want to look at the following quantity 
\begin{equation}
\label{eq:xi_convenient}
\xi(s) \equiv \frac{m^{-2}\mathcal{F}^\Theta_{2,0}(s)+2}{\bar\lambda/4\pi}.
\end{equation}
 In Fig. \ref{fig:RationalApprox1}, we show $\xi(s)$ computed using LCT with $\Delta_{\rm max}=40$ at leading order in $\bar\lambda$ and compare it with the exact one-loop form factor  \eqref{eq:FF_perturbative}. In the left plot we show the ``raw'' LCT result obtained with \eqref{eq:TruncFormFactorBasic}, while in the right plot we show the improved form factor from fitting to a rational function of the form (\ref{eq:RationalApproxAnsatz}) with $N=3$.
The residuals are shown in the insets. They  are small in both cases, but fitting to the rational Ansatz improves the errors by removing the ``high-frequency'' components and reducing the overall error by about two orders of magnitude.  Moreover, because the Ansatz is manifestly an analytic function of $s$ (up to branch cuts and poles, by construction), we can evaluate it in the main region of interest, at $s>4m^2$.
In Fig. \ref{fig:RationalApprox2}, we compare the real and imaginary parts of the approximate one-loop form factor to its exact behavior in this regime.

In principle, if one knew the best values of the parameters in the rational function Ansatz (\ref{eq:RationalApproxAnsatz}), one could approximate the form factor arbitrarily well by increasing the number of terms $2N$ in it.  In any practical computation, we know at most a finite number of these terms, but the Ansatz still approaches the correct form factor fairly quickly as the number of such terms increases. 
The  basic reason  is that the form factor has a branch cut at $4m^2 < s < \infty$, and the $\newRho$ variable maps this branch cut to the boundary of the unit disk.  We notice that for functions with this behavior, rational approximations converge much better near the branch cut than do series expansions, which require a large number of terms to   approximate the behavior near the branch cut.

It is interesting and encouraging that the simple rational function approximation \eqref{eq:RationalApproxAnsatz} provides a fairly simple way to continue the ``raw'' LCT result for the form factor at $s<0$ into the region of interest $s>4m^2$.  The main advantage of this method is that it is easy to implement.  On the other hand, its main disadvantage is that it is not as systematic as we would like.  In particular, the accuracy of the result at $s>4m^2$ depends on the order $N$ of the rational function being used, and even at the largest values of $\Delta_{\rm max}$ that we could run, the result is more sensitive to $N$ than we would like (for instance, the accuracy tends to degrade for $N \ge 5$).  We have not used the fact that the absolute value of the form factor at $4m^2 < s < 16m^2$ is known from the LCT computation of the spectral density (see section \ref{sec:SDandPade}), and perhaps this information could be used to make the rational function approximation more robust.  While we think that there is potential for improvement along these lines, we will not pursue this direction further in this paper.  Instead, we will focus in the next subsection on improving the result for the form factor in the regime $s<0$, so that it is as accurate as possible.

\subsubsection{Rational Approximation from Analyticity}
\label{sec:rational_approximation_analyticity}

We are not quite done improving the accuracy of our form factor from truncation.  To go farther, we need to understand in more detail why the ``raw'' LCT result, which represents the form factor as a sum over polynomials in the momentum ratio $X$, is quite accurate at $0<X<1$ (i.e. $s<0$), but completely incorrect anywhere else in the complex plane.  As we have discussed, the analytic structure of the full form factor is that it has a branch cut along the ray $X<0$ but otherwise is analytic; by contrast, polynomials are analytic everywhere except at $\infty$.  How then can there be any sense in which the infinite sum over polynomials converges to the correct analytic function everywhere in the complex plane?

\begin{figure}[t]
\begin{center}
\includegraphics[width=0.4\textwidth]{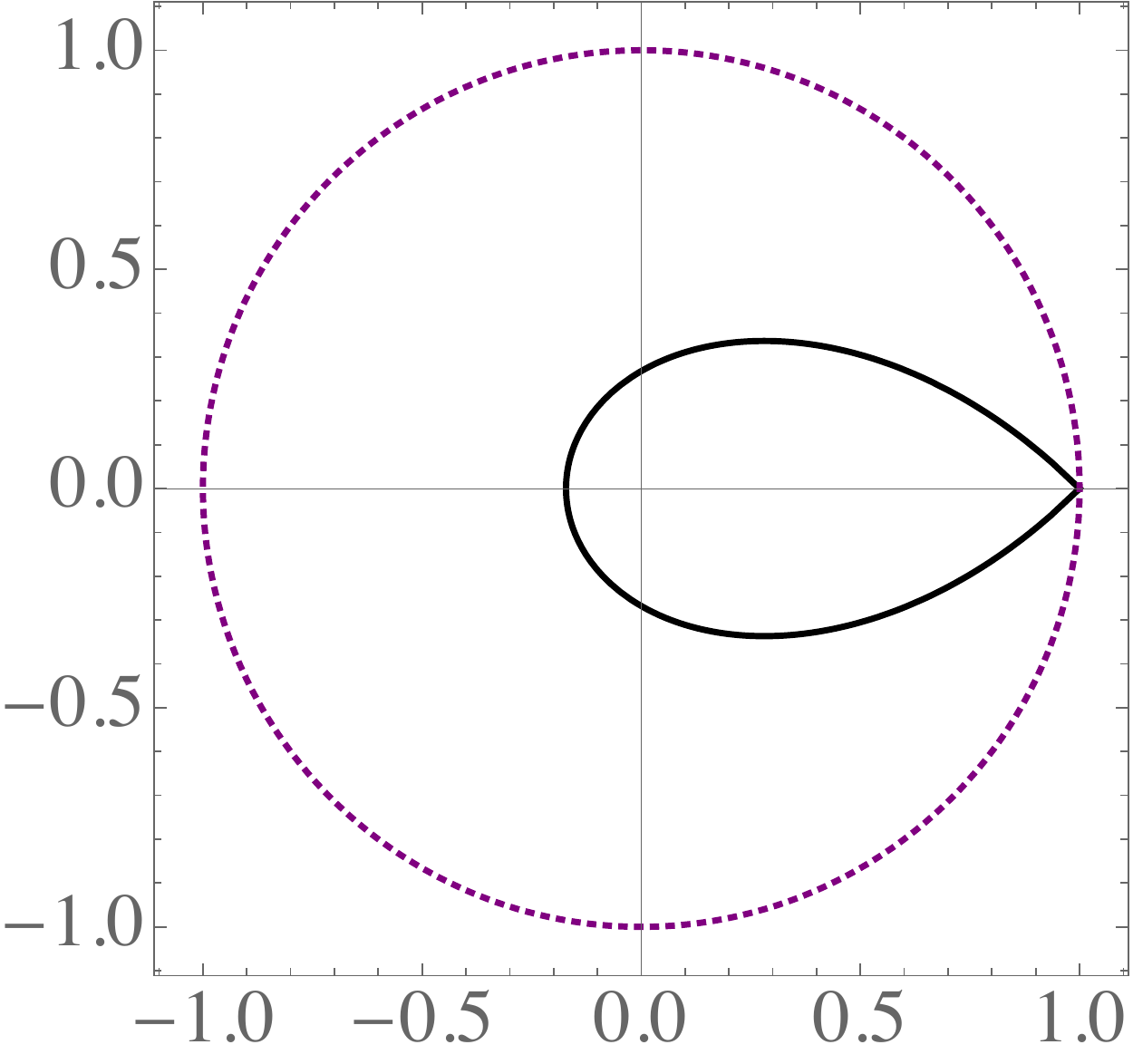}
\caption{Contour $|w|=1$ (black, solid) shown in the $\rho$ plane.  The branch cut at $w>1$ is mapped to $|\rho|=1$ (purple, dotted), so that all points along the $|w|=1$ contour except for $w=1$ are at finite distance from the radius of convergence.}
\label{fig:WVsRho}
\end{center}
\end{figure}

To see how we might make sense of this problem, it helps to consider a similar but technically simpler problem, of the decomposition of the function $\log x$ into a sum over Chebyshev polynomials $T_n$:
\begin{equation}
\log x = -\log 4 + \sum_{n=1}^\infty \frac{2}{n} T_n(1-2x) .
\end{equation}
Like our form factor decomposition, the above decomposition converges for $0<x<1$ but diverges everywhere else in the complex plane. Nevertheless, we can put it into a more favorable form by changing variables:
\begin{equation}
1-2x = \frac{w+w^{-1}}{2} \quad\Rightarrow\quad T_n(1-2x) = \frac{w^n+w^{-n}}{2}.
\end{equation}
Collecting the positive powers of $w$ into a function $f(w)$ and the negative powers of $w$ into a function $f(w^{-1})$, we can rewrite the decomposition of $\log x$ as
\begin{equation}
\log x = f(w) + f(w^{-1}), \qquad f(w) \equiv - \log 2 + \sum_{n=1}^\infty \frac{w^n}{n} = \log(\frac{1-w}{2}).
\end{equation}
Now it is easy to see that indeed $f(w) + f(w^{-1}) = \log(\frac{2-w-w^{-1}}{4}) = \log x$. More importantly, however, while the partial sums in $f(w)+ f(w^{-1})$ converge only on the line segment $0<x<1$, the partial sums for $f(w)$ alone converges for $|w| \le 1$.  From this point of view, the problem with our decomposition is that we are adding up two functions ($f(w)$ and $f(w^{-1})$) that individually converge on $|w|\le 1$ and $|w|\ge 1$, so that even slight deviations from the contour $|w|=1$ push us outside one or the other domain of convergence.  However, since we know that the function $f(w)$ is analytic except on the ray $w>1$, we can improve its convergence by mapping to the appropriate variable
$\rho$ related to $w$ as
\begin{equation}
\label{eq:rho_coordinate}
w=\frac{4 \rho}{(1+\rho)^2}.
\end{equation}
This map pushes the contour $|w|=1$, where we want to evaluate $f(w)$, deeper inside a ball where its Taylor series (now in $\rho$) converges, as depicted in Fig. \ref{fig:WVsRho}.\footnote{In general, $|\rho(w)| = |\rho(w^{-1})|$ for $|w|=1$, and moving away from $|w|=1$ decreases one of $|\rho(w)|$ or $|\rho(w^{-1})|$ at the expense of increasing the other, so the accuracy of this method will still be best on the contour $|w|=1$.}

This example illustrates how we can improve the result for the form factors. From the expression in \eqref{eq:final_raw_from_FF}, we can transform to the $w$ coordinate, take the positive power terms and then transform to the $\rho$ coordinate. Finally, we perform a Pad\'e approximation in the $\rho$ variable, and add the corresponding contribution from the negative $w$ power terms to get the form factor.

\subsection{Final Results for the Form Factors at $s<0$}
\label{sec:FF_final_results}
In the last subsection, we discussed two procedures one can do to the "raw" form factors from LCT to reduce the uncertainties. The fist one is to use a simple rational function fit to the "raw" form factor, while the second one is more complicated, which involves a Pad\'e approximation after two coordinate transformations. The second procedure makes use of the analytic properties of the form factor, and is supposed to give better result. Here, we present the result obtained through the second procedure. 

To summarize, to get an accurate result for the form factor, we first compute the form factor from LCT as a sum over Jacobi polynomials given by (\ref{eq:TruncFormFactorBasic}), then we transforms all these different Jacobi polynomials to the ones of the form $\tilde{P}_{n}^{(1,-2)}(1-2X)$, such that the form factor is brought to the form \eqref{eq:final_raw_from_FF}, which we write here again for convenience
\begin{equation}
m^{-2}\CF_{1,1}^\Theta(X)+2 =
\sum_{n} \widetilde{a}_{n} \widetilde{P}_{n}^{(1,-2)}(1-2 X),
\end{equation}
We remind that the relation between the $s$ and $X$ variables is given by \eqref{eq:s_X_rel}.
We compute ${\tilde{a}_n}$ coefficients at different $\Delta_\text{max}$, up to $\Delta_\text{max}=40$, and extrapolate them to $\Delta_\text{max}=\infty$, and only keep those coefficients that have converged (in the results presented in this paper, we keep $\tilde{a}_n$ up to $n=18$). We then transform the result to the $w$ coordinate related to X by
\begin{equation}
X= \frac{1}{2} \left(1-\frac{1}{2} \left(w+\frac{1}{w}\right)\right).
\end{equation}
Then we take the positive power terms, transform to the $\rho$ coordinate by \eqref{eq:rho_coordinate}  and perform a Pad\'e approximation in $\rho$ at $\rho=0$. Finally, we take the sum of this Pad\'e approximation and the corresponding one from negative $w$ power terms, and use the relationship between $\rho$ and $s$ to write the result as a function of $s$. 

\begin{figure}[t]
\begin{center}
\includegraphics[width=\textwidth]{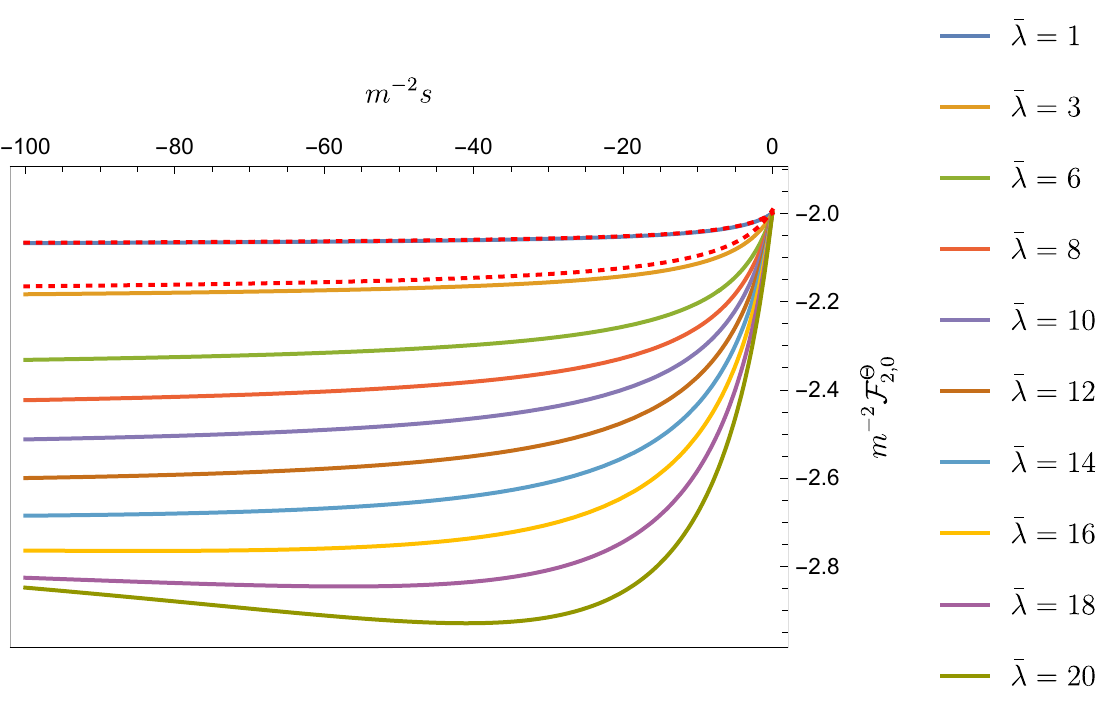}
\caption{The form factor of the trace of the stress tensor  $\CF_{2,0}^\Theta$ in the $\phi^4$ model computed for various values of $\bar\lambda$ from LCT. As a comparison, we also plotted  the perturbative two-loop form factor (equation (\ref{eq:FF_pert})) for $\bar\lambda=1$ and $\bar\lambda=3$ (red dotted lines). One can see that for $\bar\lambda=1$, the perturbative result  agrees with the LCT non-perturbative $\bar\lambda=1$ result very well as expected, since $\bar\lambda=1$ is in the perturbative regime ($\bar\lambda/4\pi\ll 1$), while for $\bar\lambda=3$, they start to deviate. This provides a consistency check for our numerical code for computing the non-perturbative form factors from LCT.}
\label{fig:FFplot}
\end{center}
\end{figure}

The final result for the form factor is shown in fig. \ref{fig:FFplot}, where we also show that in the perturbative regime, the form factor obtained this way agrees with analytic perturbative two-loop form factor (equation (\ref{eq:FF_pert})) very well. The uncertainty in the form factor increases as we increases $\bar\lambda$, due the slower convergence rate of the coefficients $\tilde a_n$ at larger $\bar \lambda$. The plots in fig. \ref{fig:FFplot} are obtained from degree $(7,7)$ Pad\'e approximant. To roughly estimate the uncertainty of the final form factors we got, we compared the the degree $(7,7)$ Pad\'e approximation with those of degree $(7,8)$, $(8,7)$ and $(8,8)$, and the differences between them range from $\CO(10^{-5})$ at $\bar\lambda=1$ to $\CO(10^{-3})$ at $\bar\lambda=12$, and $\CO(10^{-2})$ at $\bar\lambda=20$ (this is at $s\sim-100m^2$, and the convergence is much better for smaller $|s|$.).

\subsection{Analytic Continuation to $s>0$}
\label{sec:FFPositives}
\begin{figure}[h]
\begin{center}
\includegraphics[width=0.9\textwidth]{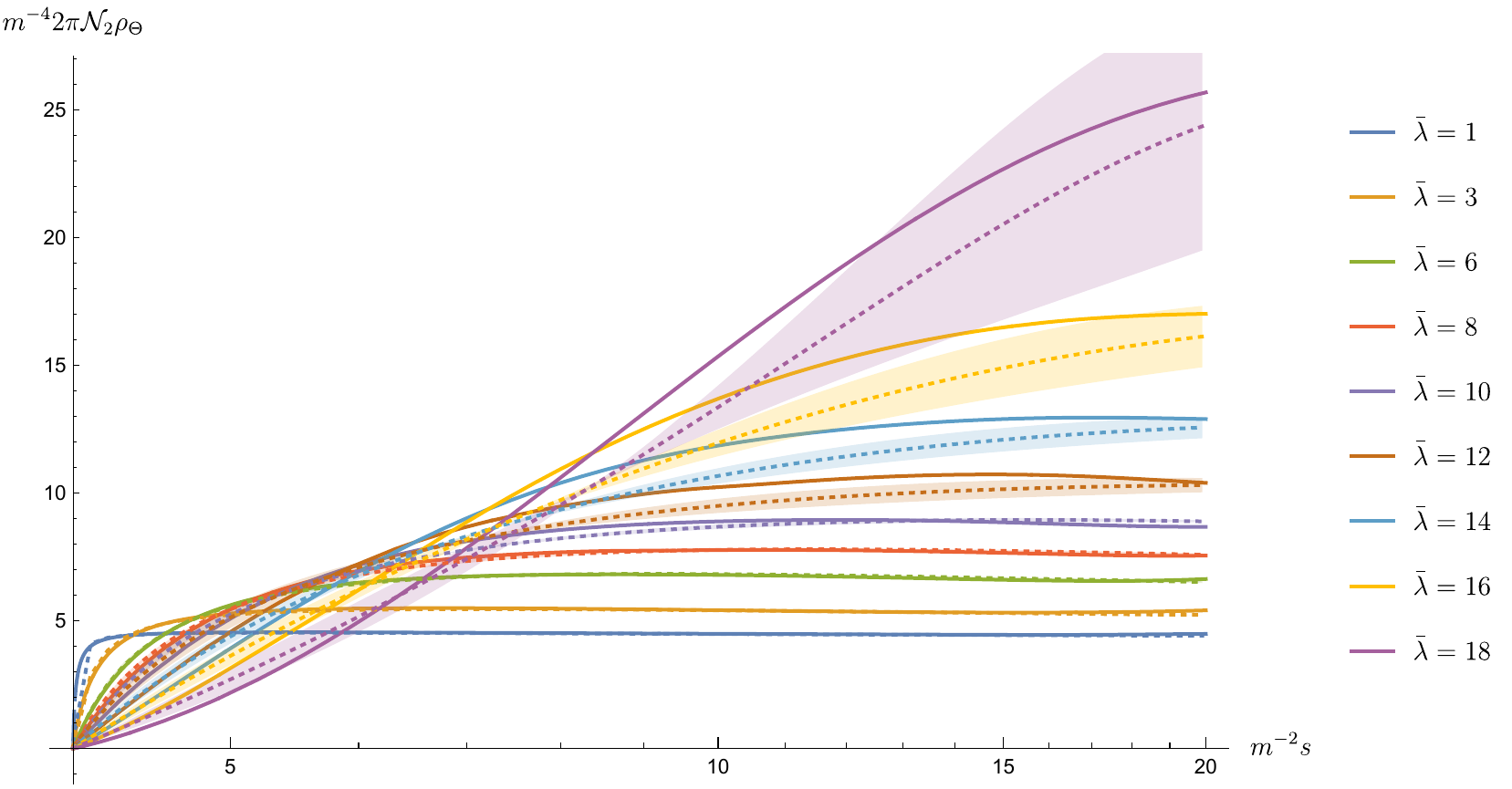}
\caption{Comparison of $2\pi\mathcal{N}_2\rho_\Theta(s)$ computed in section \ref{sec:SD_final_results}  (solid lines) and reconstructed via \eqref{eq:elastic_FF_SD} from the analytically continued form factor obtained in this section  (dotted lines).
The dotted lines are the average of the rational fits with degree $N=3,4$, and $5$, while the shaded areas are the uncertainties determined by the differences in these three rational fits.}
\label{fig:ffvsSD}
\end{center}
\end{figure}

\begin{figure}[h!]
\begin{center}
\includegraphics[width=0.45\textwidth]{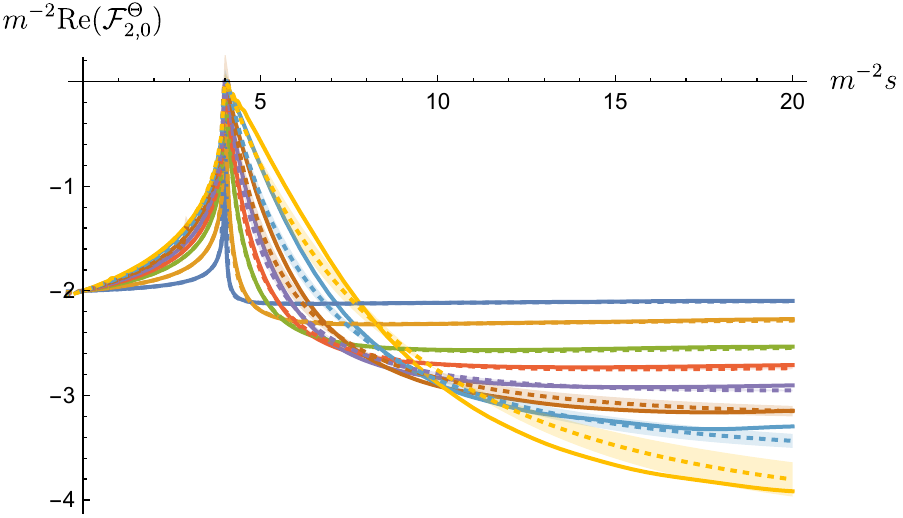}\quad
\includegraphics[width=0.5\textwidth]{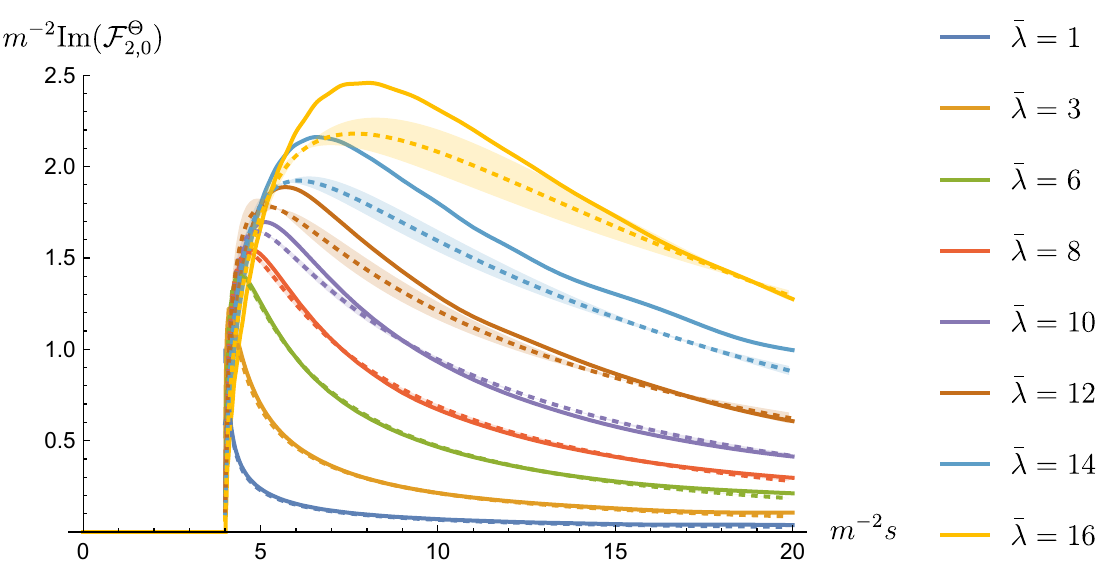}
\caption{Form factors in the $s>0$ regime. The solid lines are obtained through the S-matrix/form factor bootstrap developed in \cite{truncboot}, while the dotted lines are the average of rational approximations with degrees $N=3,4$ and $5$, and the shaded areas are the uncertainties determined by the difference in these three rational approximations. }
\label{fig:ffpositives}
\end{center}
\end{figure}

In the previous section, we made use of analyticity to obtain accurate results for the form factors in the $s<0$ regime. One natural question to ask is: if we simply analytically continue these form factors to get the results at $s>0$, how good are they? In section \ref{sec:FFRationalFit}, we studied this question for the one-loop form factor by analytically continuing the rational approximation of the form factor at $s<0$ directly from LCT, and the result in figure \ref{fig:RationalApprox2} looks reasonably good. In this section, we revisit this problem by analytically continuing the non-perturbative results obtained in section \ref{sec:FF_final_results}.
In fact, we need to perform an rational approximation of the form (\ref{eq:RationalApproxAnsatz}) to the result obtained in last subsection before we analytically continue, since the result of last subsection was optimized specially for $s<0$, and the analytic structure of it is actually not right at $s>0$. We show the results in figure \ref{fig:ffvsSD} and \ref{fig:ffpositives}.

The above method of analytic continuation is not rigorous and one can ask a question: how trustworthy these results are? In the ``elastic'' regime it is easy to answer this question by recalling that the spectral density is related to the two-particle form factor as
\begin{equation}
\label{eq:elastic_FF_SD}
\rho_\Theta(s) =
(2\pi \mathcal{N}_2)^{-1}|\mathcal{F}^\Theta_{2,0}(s)|^2,\qquad
s\in[4m^2,16m^2].
\end{equation}
In the ``elastic'' regime we can thus reconstruct the spectral density from the obtained analytic continuation of the LCT form factors and compare it with the direct LCT results for the spectral density obtained in section \ref{sec:SD_final_results}. We show this comparison in figure \ref{fig:ffvsSD}. One sees a good agreement for lower values of $\bar\lambda$ and less perfect agreement for higher values $\bar\lambda$. Nevertheless even for larger values of $\bar\lambda$ the rough shape of the spectral density reconstructed from the analytically continued LCT form factors is roughly the same as the direct computed LCT spectral density.

In the companion paper \cite{truncboot} we will obtain the two-particle form factor more rigorously by combining the LCT data with the S-matrix/form factor bootstrap. For completeness we show in figure \ref{fig:ffpositives} the form factors obtained in \cite{truncboot} compared with the less rigorous ones obtained here. As already expected from figure \ref{fig:ffvsSD} there is a very good agreement for smaller values $\bar\lambda$ and less perfect agreement for larger values of $\bar\lambda$.

\newpage
\section{Discussion}
\label{sec:discussion}

The main goal of this paper was to develop new methods to obtain numerical nonperturbative results in LCT. We focused on the spectral density and the two-particle form factors of the stress tensor in the 2d $\phi^4$ theory.  While we only studied the case of the stress tensor and $d=2$, the idea behind this construction was fairly general and should be applicable to any UV CFT operator in any theory to which LCT can be applied, regardless of spacetime dimension.\footnote{See e.g. \cite{Hogervorst:2014rta,Katz:2016hxp,Elias-Miro:2020qwz,Anand:2020qnp} for recent equal-time and lightcone truncation studies of $\phi^4$ in $d>2$, and in particular the recent progress in \cite{Elias-Miro:2020qwz,Anand:2020qnp} on understanding how to handle state-dependent counter-terms in $d=3$.}  There are many other interesting CFTs to which these methods can be applied, and in particular another natural application where there has been significant lightcone Hamiltonian truncation work is QCD in two dimensions \cite{Dempsey:2021xpf,Katz:2014uoa,Katz:2013qua,Bhanot:1993xp,Demeterfi:1993rs,dalley1993string}.  Moreover, we think it is likely that one could generalize this approach to equal-time truncation computations as well without much trouble.  The main advantage of lightcone truncation was that the Hamiltonian preserves boosts, so obtaining one-particle states as a function of their momentum is trivial.  However, in equal-time truncation, one may be able to simply repeat the diagonalization procedure in many different boosted frames.  It would be interesting to develop such an approach in an equal-time setting.\footnote{See also \cite{Pozsgay:2007kn,acerbi1996form} for another approach to computing form factors using TCSA.}  In principle, the spectral densities should also be computable in equal-time truncation, so all of the LCT data obtained in this work could be reproduced with an equal-time approach, which would provide a useful check. 

Perhaps most generally, we hope that this work might help to encourage further efforts among future Hamiltonian truncation studies of all stripes  to obtain a wide range of dynamical observables such as form factors and spectral densities, and beyond.

\begin{center}
\subsection*{Acknowledgments}
\end{center}

We thank Ami Katz, Matthew Walters, for helpful conversations and comments on a draft. ALF and HC were supported in part by the US Department of Energy Office of Science under Award Number DE-SC0015845 and the Simons Collaboration Grant on the Non-Perturbative Bootstrap, and ALF in part by a Sloan Foundation fellowship. 

\

\appendix

\section{Summary of Conventions}
\label{app:definitions}

In this work we use the ``mostly plus'' Lorentzian metric, $\eta^{\mu\nu} = {\rm diag} (-1, +1, +1, \dots)$. Our methods can be applied to form factors and spectral densities of general local operators, but we will mostly focus on the stress tensor for concreteness. We will follow convention and define $\Theta$ to be the trace of the stress tensor,
\begin{equation}
\label{eq:trace}
\Theta(x) \equiv \eta_{\mu\nu} T^{\mu\nu}(x).
\end{equation}
When we compute two-point functions, we use subscripts to distinguish the Wightman two-point function,
\begin{equation}
\< {\rm vac} |\Theta(x_1)\Theta(x_2)|{\rm vac}\>_W \equiv \lim_{\epsilon\rightarrow 0^+} \< {\rm vac} |\Theta(x_1^0-i\epsilon, \vec x_1)\Theta(x_2)|{\rm vac}\>
\end{equation}
form the time-ordered two-point function
\begin{equation}
\< {\rm vac} |\Theta(x_1)\Theta(x_2)|{\rm vac}\>_T \equiv
\theta(x_1^0-x_2^0 ) \< {\rm vac} |\Theta(x_1)\Theta(x_2)|{\rm vac}\>_W +
\theta(x_2^0-x_1^0 ) \< {\rm vac} |\Theta(x_2)\Theta(x_1)|{\rm vac}\>_W.
\end{equation}
The spectral density $\rho_{\Theta}$ is related to these two-point functions as follows \cite{Weinberg:1995mt}:
\begin{equation}
2\pi\theta(p^0)\rho_\Theta(-p^2) \equiv \int d^dx e^{-i p\cdot x}
\< {\rm vac} |\Theta(x)\Theta(0)|{\rm vac}\>_W,
\end{equation}
\begin{equation}
\label{eq:SD_TimeOrderedCorrelator}
2\pi\theta(p^0)\rho_\Theta(-p^2) = 2\,\text{Re}
\int d^dx e^{-i p\cdot x}
\< {\rm vac} |\Theta(x)\Theta(0)|{\rm vac}\>_T.
\end{equation}

We denote the one-particle stable states with mass $m$ and momentum $\vec p$ as
\begin{equation}
\label{eq:1PS}
|m,\vec p\,\>,
\end{equation}
with the conventional  normalization,
\begin{equation}
\label{eq:normalization_1PS}
\<m,\vec p_1\,|m,\vec p_2\,\> = 
2\omega_{p_1} (2\pi)^{d-1}\delta^{d-1}(\vec p_1-\vec p_2) = 2 p_{1-} (2\pi)^{d-1} \delta(p_{1-}-p_{2-}) \delta^{(d-2)}(p_{1\perp} -p_{2\perp}),
\end{equation}
where $\omega_{p_1}^2 \equiv \vec{p}_1{}^2 + m^2$,  $p_{\pm} \equiv \frac{\omega_p \pm p_x}{\sqrt{2}}$ are lightfront momentum, and $p_{\perp}$ is the momentum perpendicular to the lightfront.

We denote the two-particle in and out asymptotic states by 
\begin{equation}
\label{eq:2PS}
|m,\vec p_1;m,\vec p_2\,\>_\text{in}
\quad\text{and}\quad
|m,\vec p_1;m,\vec p_2\,\>_\text{out}.
\end{equation}
with the following normalization:
\begin{multline}
\label{eq:normalization_2PS}
{}_\text{in}\<m,\vec k_1;m,\vec k_2|m,\vec p_1;m,\vec p_2\,\>_\text{in} =
{}_\text{out}\<m,\vec k_1;m,\vec k_2|m,\vec p_1;m,\vec p_2\,\>_\text{out} =\\
4\sqrt{m^2+\vec p_1^{\,2}}\sqrt{m^2+\vec p_2^{\,2}}\, (2\pi)^{2(d-1)}\delta^{(d-1)}(\vec p_1-\vec k_1)\delta^{(d-1)}(\vec p_2-\vec k_2)+
(\vec p_1 \leftrightarrow \vec p_2).
\end{multline}
Form factors are defined as matrix elements of local operators in the basis of asymptotic states.  We will use subscripts to denote the number of `in' and `out' particles in the ket and bra of these matrix elements, e.g.
\begin{equation}
\label{eq:form_factor}
\begin{aligned}
\mathcal{F}^\Theta_{1,1}(t)&\equiv
{}_\text{out}\<m,\vec p_1\,|\Theta(0)|m,\vec p_2\>_\text{in},\\
\mathcal{F}^\Theta_{2,0}(s)&\equiv
{}_\text{out}\<m,\vec p_1\,;m,\vec p_2\,|\Theta(0)|{\rm vac}\>.
\end{aligned}
\end{equation}
The analogues of the Mandelstam variables for the form factors  are defined as
\begin{equation}
\label{eq:mandelstam_analogue}
s \equiv  -(p_1+p_2)^2,\qquad
t \equiv  -(p_1-p_2)^2 = 4m^2-s.
\end{equation}
In the case of a scalar operator, such as $\Theta$, the form factors depend only on the Mandelstam variable $s$.  Moreover, in this case $\mathcal{F}_{2,0}$ and $\mathcal{F}_{1,1}$ are related by crossing symmetry: 
\begin{equation}
\label{eq:relation_FF}
\mathcal{F}^\Theta_{2,0}(s) = \mathcal{F}^\Theta_{1,1}(s) ,
\end{equation}
that is, they are (analytic continuations of) the same function (note that $s$ and $t$ were swapped in their definition (\ref{eq:form_factor})).
At $p_1=p_2$, $\mathcal{F}_{1,1}$ is the diagonal matrix element of $\Theta$ in the one-particle state $|m, \vec{p}_1\>$.  Moreover, no momentum flows through $\Theta$ in this case, so this matrix element just measures the rest mass of the particle. As a consequence, 
we have the following exact condition:\footnote{See e.g. appendix G in \cite{Karateev:2020axc} for details.}
\begin{equation}
\label{eq:mass_definition}
\lim_{s\rightarrow 0} \mathcal{F}^\Theta_{2,0}(s) = -2m^2.
\end{equation}

In $d=2$, the $C$-function \cite{Zamolodchikov:1986gt,Cardy:1988tj} is directly related to the $\Theta$ spectral density:
\begin{equation}
\label{eq:C-function}
C(s) \equiv 12\pi\int_{0}^{s'} ds'\, \frac{\rho_\Theta(s')}{s^{'2}}.
\end{equation}
The central charge of the UV CFT is simply $c_{\rm UV} = C(\infty)$. 
By inserting a complete set of states in the form of asymptotic states, the spectral density can be written as a sum over contributions  with definite particle number $n$:
\begin{equation}
\label{eq:SD_FF}
\rho_\Theta(s) =\sum_{n=1}^\infty \rho^{(n)}_\Theta(s) \theta(s-n^2 m^2),
\end{equation}
where the superscript $(n)$ denotes the $n$-particle part of the spectral density.  
 In particular, the two-particle part of the spectral density is related to the two-particle form factor as
\begin{equation}
\label{eq:SD_FF_rel}
\rho^{(2)}_\Theta(s) =
(2\pi \mathcal{N}_2)^{-1}|\mathcal{F}^\Theta_{2,0}(s)|^2.
\end{equation}

\section{Two-point Functions at Large Energy}
\label{app:large_energy}
Consider the time-ordered two-point correlator of the stress tensor. Expanding its spectral representation \eqref{eq:TOCandSD} in $1/s$ series we get
\be
\label{eq:LargeSExpansionTOC}
\mathbf{\Delta}_{T_{--}}(p) = i \sum_{n=0}^\infty s^{-n-1}  \sum_i | \< T_{--}(0) | \mu_i^2, p\> |^2 \mu_i^{2n} =i \sum_{n=0}^\infty s^{-n-1} \< T_{--} |M^{2n} | T_{--} \>,
\ee
where we have used the fact that the states $|\mu_i^2,p\>$ are eigenvalues of the mass-squared operator $M^2= 2 p_- \int_{-\infty}^\infty dx^- V(x)$.  For $n\ge 2$, the operator product $(M^2)^n$ is typically singular and must be regulated, which moreover introduces additional $\log s$ dependence. However, the special cases $n=0$ and $n=1$ are readily evaluated.  In fact, because $|T_{--}\>$ is already one of the states in the LCT basis, $|T_{--} \> = \frac{1}{\sqrt{12\pi}} | (\partial_- \phi)^2; p \>$, the  $n=0$ term is manifestly\footnote{The state $|T_{--}\>$ in equation (\ref{eq:LargeSExpansionTOC}) is slightly schematic; an overall normalization, including a momentum-conserving $\delta$ function, has been implicitly factored out, so that $\< T_{--} | T_{--}\> \rightarrow \frac{1}{12\pi}$.}
\be
\mathbf{\Delta}_{T_{--}}(p) = \frac{i}{12\pi s} + \dots
\ee
This agrees with the fact that the coefficient of $1/s$ in the $T_{--}$ time-ordered two-point function is fixed by the UV central charge:
\begin{align}
\nn
\mathbf{\Delta}_{T_{--}}(p) &= \int_0^\infty d\mu^2 \rho_{T_{--}}(\mu^2) \frac{i}{s-\mu^2 + i \epsilon}
\\
& =  \frac{i}{s} \int_0^\infty d\mu^2 \rho_{T_{--}}(\mu^2) + \CO(s^{-2}) = \frac{i c_{\rm UV}}{12 \pi s} + \CO(s^{-2}).
\label{eq:LargeSSD}
\end{align}
The next term is only slightly more complicated, and requires computing a single matrix element of $M^2$:\ \footnote{See e.g. tables 8 and 10 of \cite{Anand:2020gnn}.}
\be
\< (\partial_- \phi)^2 | M^2 | (\partial_- \phi)^2 \> = 6m_0^2 + \frac{3 \lambda}{4 \pi} . 
\ee
Therefore, we find that the first two powers of $1/s$ of $\mathbf{\Delta}_{\Theta}$ are
\be
\label{eq:TOCsubleading_app}
\mathbf{\Delta}_{\Theta}(p) = s^2 \mathbf{\Delta}_{T_{--}}(p)=  \frac{is^2}{12 \pi } \left( \frac{1}{s}  + \frac{6m_0^2 + \frac{3 \lambda}{4 \pi} }{s^2} +\dots \right) .
\ee

\section{Truncation Results in Perturbation Theory} 
\label{app:perturbativeLCT}

As a check of our truncation results for the form factor and spectral density, we can use time-independent perturbation theory to isolate specific orders in perturbation theory.  For instance, if we want to compare the $\CO(\lambda)$ truncation result with the Feynman diagram result, it is much more accurate to do the truncation computation with time-independent perturbation theory than it is to compute the all-orders result at small $\lambda$ and try to numerically extract the leading linear-in-$\lambda$ dependence.  Moreover, at $\CO(\lambda^n)$, only states with at most $1+\lfloor \frac{n}{2} \rfloor$ or $2+\lfloor \frac{n}{2} \rfloor$ particles contribute in the form factor or spectral density, respectively, which allows us to go to much higher $\Delta_{\rm max}$ for low orders in $\lambda$.  

\subsection{Perturbative Results from Feynman Diagrams}

In this section we provide analytic results for the $\phi^4$ model and the 2d $O(N)$ model in the large $N$ limit from standard loop computations.
The main objects we would like to compute are the form factors of  the trace of the stress-tensor and the spectral density defined in appendix \ref{app:definitions}.
In $d=2$ all these observables are functions of a single variable $s$. 
The relation between the lightcone quantization bare mass $m_0$ and the physical mass $m$ is given by \cite{Fitzpatrick:2018xlz}
\begin{equation}
\label{eq:mass0_pert}
m = m_0\left(1 - \frac{\bar\lambda^2}{768} + O(\bar\lambda^3) \right).
\end{equation}

 The form factor up to $\CO(\bar{\lambda}^2)$ is given by the following simple  expression:
\begin{multline}
\label{eq:FF_pert}
m^{-2}\mathcal{F}^\Theta_{2,0}(s) = -2+\left(\frac{\bar\lambda}{4\pi}\right)\, \Delta(s)+\\
\frac{1}{2}\left(\frac{\bar\lambda}{4\pi}\right)^2
\left(\frac{\pi^2s}{8(s-4m^2)}-\Delta(s)\left(\Delta(s)/2+1\right)\right)+\mathcal{O}(\bar\lambda^3),
\end{multline}
where, for any complex $s$, the function $\Delta(s)$ is defined as
\begin{equation}\label{eq:DeltaDef}
\Delta(s) \equiv-1+\lim_{\epsilon\rightarrow 0^+}
\frac{4m^2\text{ArcTan}\left(\frac{\sqrt{s}}{\sqrt{4m^2-s-i\epsilon}}\right)}{\sqrt{s}\sqrt{4m^2-s-i\epsilon}} .
\end{equation}

The  spectral density of the trace of the stress tensor up to $\CO(\bar{\lambda})$ is given by
\begin{equation}
\label{eq:SD_pert}
\frac{2\pi\mathcal{N}_2}{4m^4}\times\rho_\Theta(s) = 1 +
\frac{\bar\lambda}{4\pi}\times\left(
1+4m^2\mathcal{N}_2^{-1}
\log\left(\frac{\sqrt{s}+\sqrt{s-4m^2}}{\sqrt{s}-\sqrt{s-4m^2}}\right)
\right)
+ \mathcal{O}(\bar\lambda^2),
\end{equation}
where we have defined
\begin{equation}
\mathcal{N}_2\equiv 2\sqrt{s}\sqrt{s-4m^2}.
\end{equation}

We also reproduce here the stress tensor two-particle form factor and spectral density of the large $N$ limit ($N\rightarrow \infty $) of the 2d $O(N)$ model with the following Lagrangian:
\begin{equation}
V_{\phi^4}^{O(N)}(\phi) \equiv \frac{1}{2}m_0^2 (\phi_i\phi_i )+ \frac{\lambda}{8N} (\phi_i\phi_i) (\phi_j\phi_j).
\end{equation}
Repeated indices are summed over. This model is solvable at infinite $N$, which will provide a useful example to test some of our methods. The exact two-particle form factor and spectral density are
\begin{equation}
\label{eq:FF_largeN}
m^{-2}\mathcal{F}^\Theta_{2,0}(s) = -2+ \frac{2\bar\lambda \Delta(s)}{8\pi+\bar\lambda\,(1+\Delta(s))}.
\end{equation}
\begin{equation}
2\pi \mathcal{N}_2\rho_\Theta(s) =
|\mathcal{F}^\Theta_{2,0}(s)|^2.
\end{equation}

For details of these computations, see appendix C of \cite{truncboot}.

\subsection{Perturbative Results from LCT}
\label{app:time-independent_perturbation_theory}

\begin{figure}[t!]
\begin{center}
\includegraphics[width=0.37\textwidth]{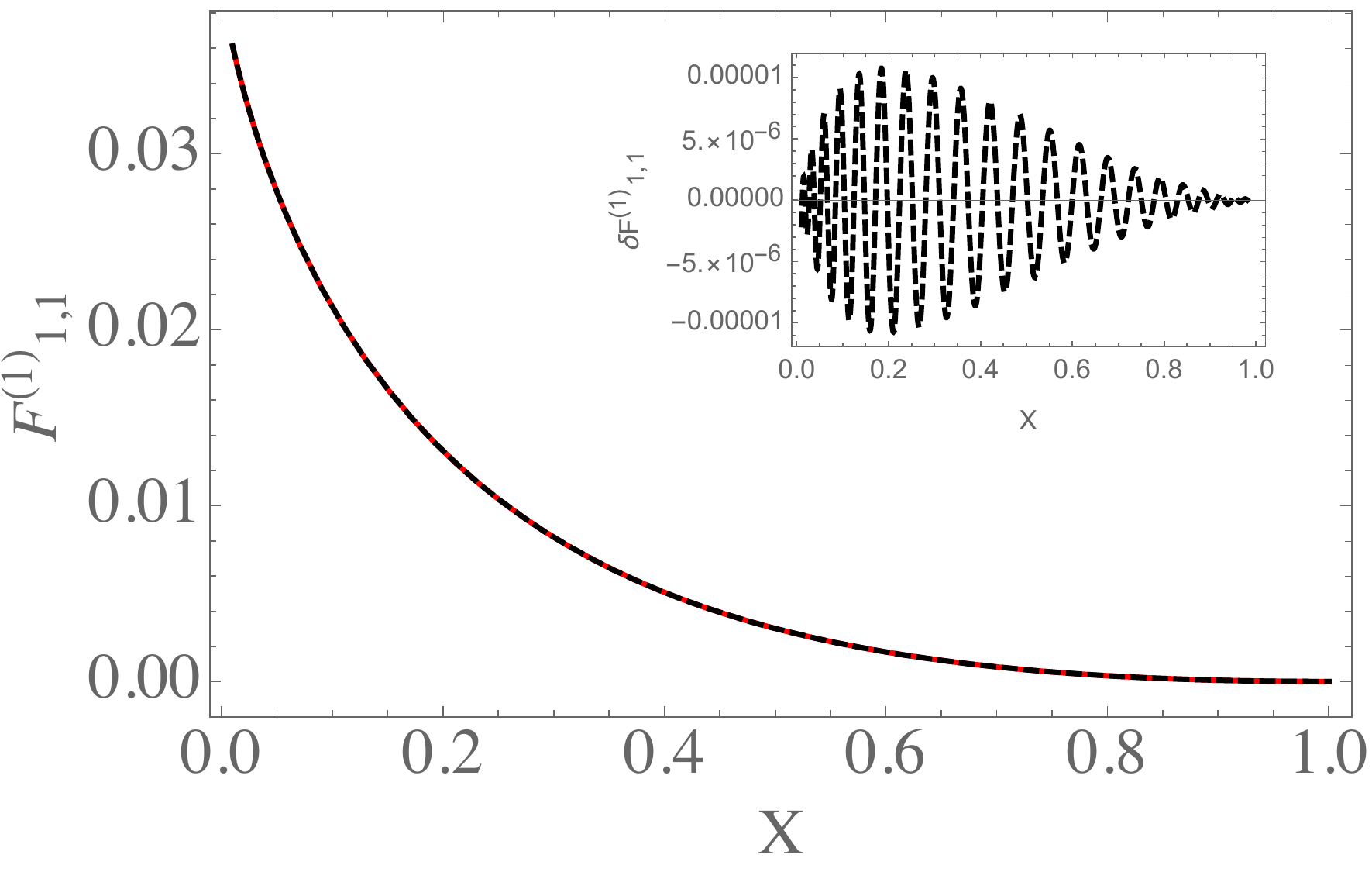}
\includegraphics[width=0.4\textwidth]{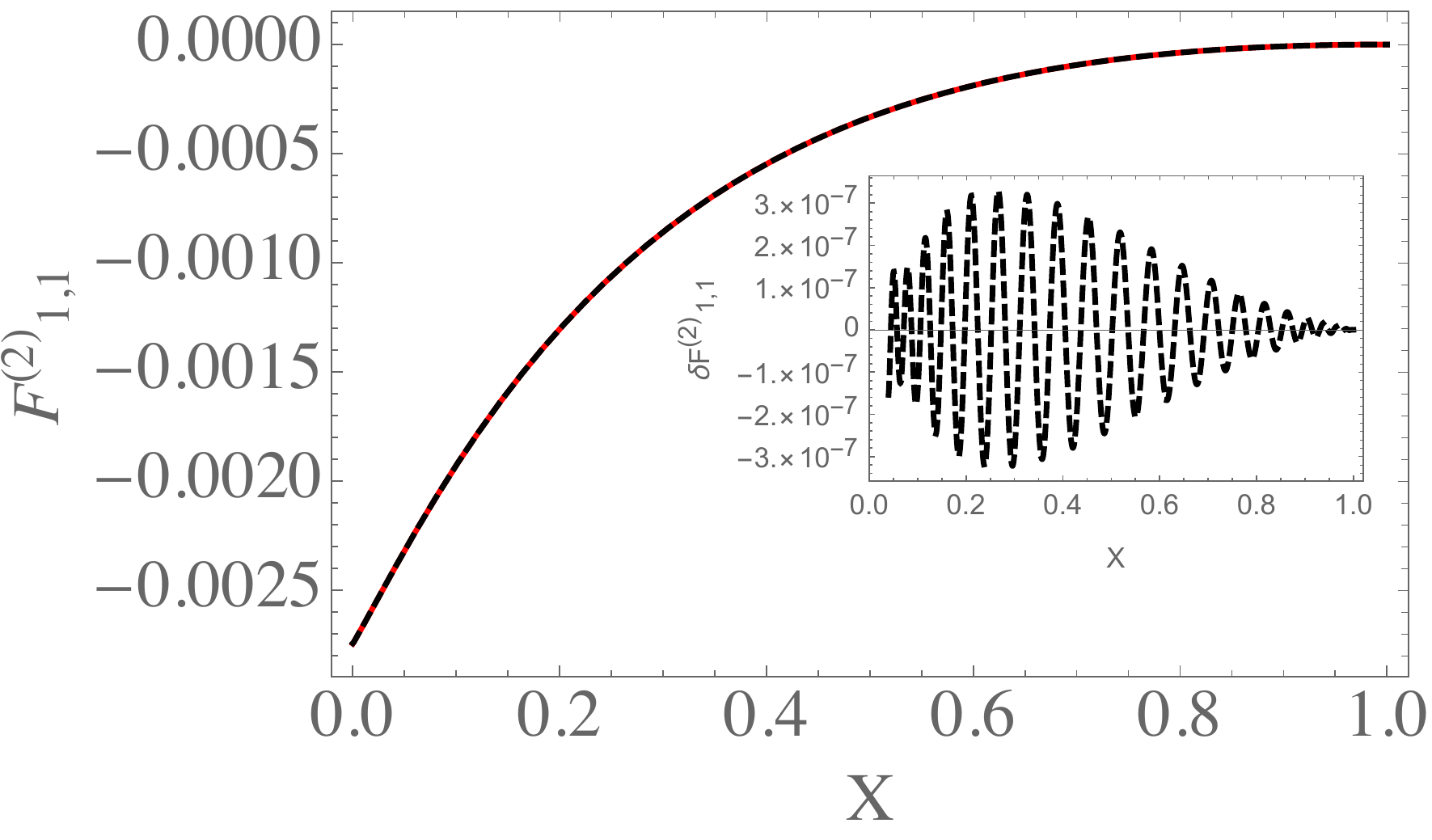}
\caption{Comparison of $\Theta$ two-particle form factor ${\cal F}^\Theta_{1,1}$ in perturbation theory from LCT ({\it black, thick}) at $\Delta_{\rm max}=50$ vs Feynman diagrams ({\it red, dashed}), as a function of $X= \frac{p_{1-}}{p_{2-}}$, for $\CO(\lambda)$ ({\it left}) and $\CO(\lambda^2)$ ({\it right}). The difference between the truncation result and the exact result is shown in the insets.}
\label{fig:PertCheck}
\end{center}
\end{figure}

To perturbatively compute the form factor in LCT, we choose a specific truncation $\Delta_{\rm max}$ and exactly diagonalize the mass term Hamiltonian $H_2 = m^2 \int dx^- \phi^2$, but treat the interaction term Hamiltonian $H_4 = 2 \frac{\lambda}{4!} \int dx^- \phi^4$ as a perturbation.  The interacting eigenstates given by the standard time-independent perturbation theory result:
\begin{eqnarray}
|n\> &=& |n^{(0)}\> + \sum_{k\ne n} \frac{V_{kn}}{E_{nk}} | k^{(0)} \> +  \sum_{k_1 \ne n} \left( \sum_{k_2 \ne n} \left( \frac{V_{k_1 k_2} V_{k_2 n}}{E_{n k_1} E_{n k_2}} \right)- \frac{ V_{nn} V_{k_1n}}{E_{n k_1}^2} \right) | k_1^{(0)} \>  \nn \\
& & -\frac{1}{2} \sum_{k_1 \ne n} \frac{V_{k_1 n} V_{n k_1}}{E_{k_1 n}^2}  |n^{(0)} \>+ \dots ,
\end{eqnarray}
where in our case the zeroth order eigenvalues $E_n$ and eigenstates $|n^{(0)}\>$ are the eigenstates of the mass term $H_2$, $E_{nk} \equiv E_n - E_k$, and $V$ is the $\phi^4$ Hamiltonian term $H_4$. The one-particle state at $\lambda=0$ is just the momentum-space state created by the primary operator $\partial \phi$.  At $\Delta_{\rm max}=5$, we worked through the $\CO(\lambda)$ form factor explicitly in section \ref{eq:PertAnalysis}.  Here we repeat the analysis, but at $\Delta_{\rm max}=50$, and we also obtain the $\CO(\lambda^2)$ piece.  Once the eigenstates are known up to a given order in $\lambda$, one simply needs to substitute their components $c_i$ in the primary operator basis into equation (\ref{eq:TruncFormFactorBasic}).  The results at $\CO(\lambda)$ and $\CO(\lambda^2)$ are shown and compared to the exact answer from (\ref{eq:FF_pert}) in Fig. \ref{fig:PertCheck}, as a function of $X$; recall that $m^{-2} s = 2 - X -X^{-1}$. 

\subsubsection*{2d $O(N)$ model in the large $N$ limit}
We also consider the infinite $N$ limit of the 2d $\CO(N)$ model.  In this case, only one- and three-particle states from the UV basis contribute to the form factor $\mathcal{F}_{1,1}^{T_{--}}$, and we can obtain analytic expressions for all the Hamiltonian matrix elements.  In fact, the only three-particle states that contribute are created by operators of the form $\partial^{k_1} \phi_i \partial^{k_2} \phi_j \partial^{k_3} \phi_j$, summed on $j$, and their wavefunctions in momentum space can be written out explicitly: 
\be
| [ \phi_i [ \phi_j \phi_j]_{\ell_1}]_{\ell_2} \> = \frac{1}{\sqrt{N}} \int dp_1 dp_2 dp_3  \delta(P-p_1-p_2-p_3) f_\ell(p_1, p_2, p_3) | p_1,i; p_2, j; p_3,j\>,
\ee
where
\be
f_\ell = (p_1+p_2)^{\ell_1} \tilde{P}^{(1,1)}_{\ell_1}(\frac{p_2-p_1}{p_1+p_2}) \tilde{P}^{(2\ell_1+3,1)}_{\ell_2}(\frac{p_3-p_1-p_2}{p_1+p_2+p_3}) .
\ee
Here, $P_n^{(a,b)}$ is a Jacobi polynomial, and the hat indicates that it is normalized,
\be
 \int_0^1 dx x^a (1-x)^b \tilde{P}_m^{(a,b)}(1-2x) \tilde{P}_n^{(a,b)}(1-2x) = 1. 
\ee
  We can choose a frame where total momentum $P=p_1+p_2 +p_3=1$, and also choose a new set of variables:
\be
p_1=x_1 x_2, \qquad p_1+p_2 = x_2
\ee
so $p_2 = x_2(1-x_1)$.

For the mass term acting on three-particle states, we have contractions where the $\phi$s from the mass term $\phi^2$ hit a $\phi_j$ from the $\phi_j \phi_j$ part of the state, and also contractions where the $\phi$s from the mass term hit a $\phi_i$ from the three-particle states.  All cases are leading order in $1/N$.  The full contribution is
\begin{eqnarray}
&& \< [\phi_i [\phi_j \phi_j]_{\ell_1} ]_{\ell_2} | \phi^2 | [ \phi_i [ \phi_j \phi_j]_{\ell_3}]_{\ell_4} \> 
 = \int dp_1 dp_2  \left( p_1 p_2 + p_1 p_3 + p_2 p_3 \right)f_{\ell}(p) f_{\ell'}(p) \\
 && \qquad  = \int x_2 dx_1 dx_2  \left( x_2(1-x_2) + x_1 (1-x_1) x_2^2\right) \nn\\
  && \qquad \times x_2^{\ell_1+ \ell_3} \tilde{P}^{(1,1)}_{\ell_1}(1-2x_1)\tilde{P}^{(1,1)}_{\ell_3}(1-2x_1)  \tilde{P}^{(2\ell_1+3,1)}_{\ell_2}(1-2x_2)\tilde{P}^{(2\ell_3+3,1)}_{\ell_4}(1-2x_2) . \nn
\end{eqnarray}
Note that this integral is a sum of two terms that each factorize into an independent $dx_1$ integral and a $dx_2$ integral.  These integrals can be evaluated efficiently by using the expressions in appendix \ref{app:jacobi} to expand the Jacobi polynomials with one index in terms of Jacobi polynomials with another index that makes them orthogonal when integrated against the appropriate measure $dx x^a (1-x)^b$.

\begin{figure}[t!]
\begin{center}
\includegraphics[width=0.48\textwidth]{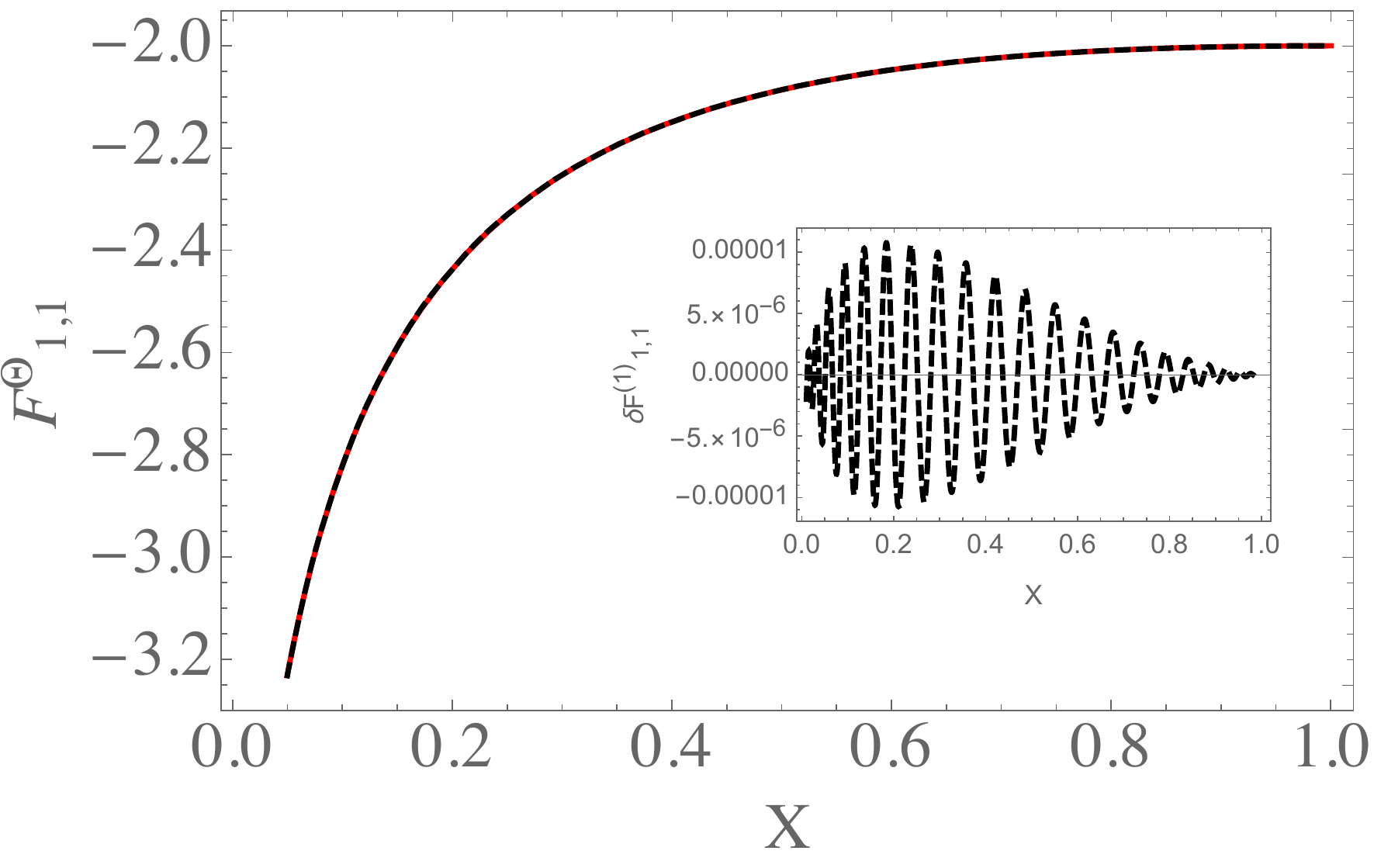}
\caption{Comparison of $\Theta$ two-particle form factor $\mathcal{F}^\Theta_{1,1}$ at infinite $N$ from LCT ({\it black, thick}) at $\Delta_{\rm max}=43$ vs resumming Feynman diagrams ({\it red, dashed}), at $\lambda=30$, as a function of $X= \frac{p_{1-}}{p_{2-}}$.  The difference between the truncation result and the exact result is shown in the inset.}
\label{fig:LargeNFFLCT}
\end{center}
\end{figure}

For the interaction itself, we have to consider 1-to-3 processes and 3-to-3 processes.  The contributions that survive at large $N$ are
\be
\< \partial \phi | \phi^4 | [ \phi_i [ \phi_j \phi_j]_{\ell_1} ]_{\ell_2} \> \propto \frac{\sqrt{N}}{32\pi} \int dp_1 dp_2 f_\ell (p)  
   =  \frac{\sqrt{N}}{8\pi} \sqrt{\frac{\left(2 l_1+3\right) \left(2 l_1+2 l_2+5\right)}{\left(l_1+1\right) \left(l_1+2\right) \left(l_2+1\right) \left(2 l_1+l_2+4\right)}}
\ee
for the 1-to-3 interactions, and
\begin{eqnarray}
&& \< [ \phi_i [\phi_j \phi_j]_{\ell_1} ]_{\ell_2} | \phi^4 | [ \phi_i [\phi_j \phi_j]_{\ell_3} ]_{\ell_4}  \> \\
   && \propto \frac{ N}{8\pi} \sqrt{ \frac{(2\ell_1+3)(2\ell_3+3)}{(1+\ell_1)(2+\ell_1) (1+\ell_3)(2+\ell_3)}} \nn\\
   && \times \int dx_2 x_2^{1+\ell_1 + \ell_3} (1-x_2)  \tilde{P}_{\ell_2}^{(2\ell_1+3, 1)}(1-2x_2)  \tilde{P}_{\ell_4}^{(2\ell_3+3, 1)}(1-2x_2)  \nn
\end{eqnarray}
   for the 3-to-3 interactions.  Finally, we need the momentum space overlaps with $T_{--}$.  At infinite $N$, the only ones that contribute are the trivial 1-to-1 matrix element, and the 1-to-3 matrix elements. The 1-to-3  momentum space overlap is
\be
\label{eq:LargeNTmmOPE}
   \< \partial \phi_i ; p | \partial_- \phi_j \partial_- \phi_j (0) | [ \phi_i [ \phi_k \phi_k]_{\ell_1} ]_{\ell_2}; p'  \> 
    \propto -\sqrt{\frac{2}{3}} \sqrt{N} \delta_{\ell_1, 0} q^3 \tilde{P}_{\ell_2}^{(3,1)}(1-2 \frac{q}{p'}) .
\ee
The infinite $N$ limit is taken so that the coupling $\lambda \sim N^{-1}$, i.e. $\lambda N$ is held fixed.  So at leading order in large $N$, the 1-to-3 interaction term in the Hamiltonian is suppressed by $N^{-1/2}$, and we just need to keep the mass term together with the 1-to-1 and 3-to-3 interaction terms.  However, we cannot simply discard the 1-to-3 interaction term.  The reason is that, although its leading effect on the energy eigenstates is $\CO(N^{-1/2})$, when we compute the form factor there is an additional $N^{1/2}$ enhancement from the overlap (\ref{eq:LargeNTmmOPE}).  Therefore our method for computing the large $N$ form factor is as follows.  We exactly diagonalize the leading order $\CO(N^0)$ Hamiltonian, at some fixed truncation $\Delta_{\rm max}$.  Then, we treat the 1-to-3 matrix elements in the Hamiltonian as a perturbative $\CO(N^{-1/2})$ interaction and compute the leading correction to the one-particle energy eigenstate using time-independent perturbation theory.  Finally, we compute the form factor by taking the overlap of this one-particle energy eigenstate computed up to $\CO(N^{-1/2})$ and looking at its overlap with the stress tensor using (\ref{eq:LargeNTmmOPE}).  The result is $\CO(N^0)$ and is the exact answer when $N=\infty$.  In Fig. \ref{fig:LargeNFFLCT}, we compare the result at $\lambda=30$ and $\Delta_{\rm max} = 43$ against the exact result from resumming Feynman diagrams.

\section{Details of OPE Coefficient Computation}
\label{app:OPE}

In this appendix, we explain our method for computing the OPE coefficients $C_{\CO jj'}$ that enter in the formula (\ref{eq:TruncFormFactorBasic}) for the form factor in LCT.  We will focus on the case $\CO= T_{--}$, our main operator of interest in this paper.

In general, if $\CO_\ell, \CO_{\ell'}$ are holomorphic primary operators, the corresponding `in' and `out' states in radial quantization can be decomposed into a sum over `monomial' operators of the form $\partial^{\bk} \phi \equiv \partial_-^{k_1} \phi \dots \partial_-^{k_n} \phi$.  So, we can begin by working out how the OPE coefficients of $T_{--} \propto (\p_- \phi)^2$ with such monomials:
\be
\< \p^{\bk} \phi | (\p_- \phi)^2(y) | \p^{\bk'} \phi \>
\ee

Matrix elements similar as this were worked out in equation (7.35) of \cite{Anand:2020gnn}, but only for the special case where the position $y$ was integrated over all space.  We can reuse the same basic approach, however. Here we will be brief, focusing on the new ingredients that are necessary for this generalization, and the interested reader should consult \cite{Anand:2020gnn} for more details.  The computation is most efficient if the states and the operator are represented in terms of the modes of $\phi$ in radial quantization of a free scalar on the cylinder, so that radial quantization creation and annihilation operators simply need to be commuted past each other.  More precisely,  the primary operators can be written in terms of monomials as
\be
| \CO\> = \sum_{\bk} C^{\CO}_{\bk, \rm RQ} | \p^{\bk} \phi\>_{\rm RQ},
\ee
where the ``Radial Quantization'' (RQ) monomial normalization is ${}_{\rm RQ}\< \p^{\bk} \phi | \p^{\bk'} \phi\>_{\rm RQ} = \delta_{\bk, \bk'}$. More explicitly,
\be
| \p^{\bk} \phi\>_{\rm RQ} = \frac{1}{\norm{\bk}} a_{\bk}^\dagger | \rm vac \>,
\ee
where $\norm{\bk} \equiv \frac{n!}{\textrm{number of permutations of } \bk}$.\footnote{More explicitly, ``number of permutations of $\bk$'' means the number of permutations of the elements of the $n$-vector $\bk$ that leave it invariant.}
The annihilation operators $a_{\bk}^\dagger$ that appear here are (products of) the `radial quantization' annihilation operators that appears in the mode expansion of $\partial \phi$ quantized on the unit circle, and should not be conflated with the annihilation operators in momentum space.  

  So, we have
\be
\< \CO_\ell | \CO(y) | \CO_{\ell'} \>  = \sum_{\bk, \bk'} C^{\ell}_{\bk, \rm RQ}C^{\ell'}_{\bk', \rm RQ} \ \ {}_{\rm RQ} \< \p^{\bk} \phi | \CO(y) | \p^{\bk'} \phi \>_{\rm RQ} 
\ee

The operator $(\partial_- \phi)^2$ contains terms of the form $\sim a^\dagger a^\dagger, a a$, and $a^\dagger a$. In the last of these three cases, the creation and annihilation operator from $(\partial_- \phi)^2$ are commuted to the left and the right, respectively, producing contributions of the form
\be
G_{\bk\bk'}^{(\p \phi)^2}(y) \cong -\frac{ 2 \CN_{\kvec} \CN_{\kvec'}}{4\pi} \sum_{\bk/k = \bk'/k'}  \sqrt{k k' }  y^{k -k' -2} \norm{\bk/k}^2 ,
\ee
where $\CN_{\kvec} \equiv \left( \frac{1}{\sqrt{4 \pi}} \right)^n\Gamma(k_1) \dots \Gamma(k_n) \sqrt{k_1 \dots k_n}.$
Because the term $\sim a^\dagger a$ does not change particle number, these terms only contribute if the number of $\phi$ particles in the `in' and `out' state are the same, i.e. if the OPE coefficient is ``$N$-to-$N$''.

By contrast, in the case of the terms of the form $\sim a^\dagger a^\dagger$, the $a^\dagger$s are both commuted all the way to the left and produce contributions of the form
\be
G_{\bk \bk'}^{(\p \phi)^2}(y) \cong-\frac{ 2 \CN_{\kvec} \CN_{\kvec'}}{4\pi} \sum_{\bk/\{k_1, k_2\} = \bk'}  \sqrt{k_1 k_2 }  y^{k_1 +k_2 -2} \norm{\bk'}^2 .
\ee
The case $\sim a a$ is similar to $a^\dagger a^\dagger$; for both $\sim a^\dagger a^\dagger$ and $\sim a a$, the $(\partial_- \phi)^2$ changes particle number by exactly two, i.e. the OPE coefficient is ``$N$-to-$N$+$2$''.
Note that because $h-k=h'-k'$ in the first case, and $h-k_1 -k_2=h'$ in the second case, in both cases the power of $y$ is
\be
y^{h-h'-h_\CO}
\ee
as it must be since $\CO_\ell, \CO_{\ell'}$ and $\CO$ are primary operators.

Combining the above expressions, we obtain the formulas we use for the OPE coefficients of $(\partial_- \phi)^2$. The $N$-to-$N$ matrix element is
\be
\< \CO_\ell | (\partial_- \phi)^2(y) | \CO_{\ell'} \>
  =  y^{\Delta_\ell-\Delta_{\ell'}-\Delta_\CO} \sum_{\bk, \bk'} C^{\ell}_{\bk, \rm RQ}C^{\ell'}_{\bk', \rm RQ}  \sum_{\bk/k = \bk'/k'} \frac{ \norm{\bk/k}^2}{\norm{\bk} \norm{\bk'}} \sqrt{k k'}
\ee
whereas for $N$-to-$N$+$2$ it is
\be
\< \CO_\ell | (\partial_- \phi)^2(y) | \CO_{\ell'} \>
  =  y^{\Delta_\ell-\Delta_{\ell'}-\Delta_\CO} \sum_{\bk, \bk'} C^{\ell}_{\bk, \rm RQ}C^{\ell'}_{\bk', \rm RQ}  \sum_{\bk= \bk'/\{k_1',k_2'\}} \frac{ \norm{\bk}}{\norm{\bk'}} \sqrt{k'_1 k'_2}
\ee

\section{Jacobi polynomial identities}
\label{app:jacobi}
In equation (\ref{eq:TruncFormFactorBasic}) of section \ref{sec:FF_LCT_subsection}, we wrote down the expression for the form factor in LCT as a sum over  Jacobi polynomials of the form $X^{h_j -1} P_{1+h_{j'} - h_j }^{(2h_j-1, -3)}(1-2X)$ (in this section, we only consider the case where $h_\CO=h_\Theta=2$ for simplicity, but the method here works for other cases as well). Here, we provide the formulas that we used to convert these different Jacobi polynomials into the form of $P_{n}^{(\alpha,-2)}(1-2 X)$, which is better for obtaining accurate result for the form factors. We first use the following recursion relation
\begin{equation}
	P_{n}^{(\alpha+1, \beta)}(x)=\frac{2}{2 n+\alpha+\beta+2} \frac{(n+\alpha+1) P_{n}^{(\alpha, \beta)}-(n+1) P_{n+1}^{(\alpha, \beta)}(x)}{1-x}
\end{equation}
$(h_j-1)$ times to turn $X^{h_{j}-1}P_{h_{j^{\prime}}-h_{j}+1}^{\left(2 h_{j}-1,-3\right)}(1-2 X)$ into a sum of $P_n^{(\alpha,-3)}(1-2X)$ with some constant coefficients, and then use the follow formula \cite{koekoek2010hypergeometric}
 \begin{equation}
P_{\ell}^{(\alpha, \beta)}(z)=\sum_{k=0}^{\ell} \mathcal{A}_{\ell k}^{(\alpha, \beta, \gamma, \delta)} P_{k}^{(\gamma, \delta)}(z), 	
 \end{equation} where
 \begin{equation}
\begin{aligned}
\mathcal{A}_{\ell k}^{(\alpha, \beta, \gamma, \delta)}=&  \frac{\Gamma(k+\gamma+\delta+1) \Gamma(\ell+k+\alpha+\beta+1) \Gamma(\ell+\alpha+1)}{\Gamma(\ell+\alpha+\beta+1) \Gamma(k+\alpha+1) \Gamma(2 k+\gamma+\delta+1) \Gamma(\ell-k+1)} \\
& \times{ }_{3}\CF_{2}(k-\ell, \ell+k+\alpha+\beta+1, k+\gamma+1 ; k+\alpha+1,2 k+\gamma+\delta+2 ; 1)
\end{aligned}	
\end{equation}
with $\beta=-3, \gamma=1$ and $\delta=-2$.

\bibliographystyle{JHEP}
\bibliography{refs}

\providecommand{\href}[2]{#2}\begingroup\raggedright\begin{thebibliography}{10}

\bibitem{Katz:2016hxp}
E.~Katz, Z.~U. Khandker and M.~T. Walters, \emph{{A Conformal Truncation
  Framework for Infinite-Volume Dynamics}},
  \href{http://dx.doi.org/10.1007/JHEP07(2016)140}{\emph{JHEP} {\bf 07} (2016)
  140}, [\href{https://arxiv.org/abs/1604.01766}{{\tt 1604.01766}}].

\bibitem{Anand:2020gnn}
N.~Anand, A.~L. Fitzpatrick, E.~Katz, Z.~U. Khandker, M.~T. Walters and Y.~Xin,
  \emph{{Introduction to Lightcone Conformal Truncation: QFT Dynamics from CFT
  Data}},  \href{https://arxiv.org/abs/2005.13544}{{\tt 2005.13544}}.

\bibitem{Burkardt}
M.~Burkardt, \emph{{Light front quantization of the Sine-Gordon model}},
  \href{http://dx.doi.org/10.1103/PhysRevD.47.4628}{\emph{Phys. Rev.} {\bf D47}
  (1993) 4628--4633}.

\bibitem{Burkardt2}
M.~Burkardt, \emph{{Much ado about nothing: Vacuum and renormalization on the
  light front}},  \href{https://arxiv.org/abs/hep-ph/9709421}{{\tt
  hep-ph/9709421}}.

\bibitem{Fitzpatrick:2018xlz}
A.~L. Fitzpatrick, E.~Katz and M.~T. Walters, \emph{{Nonperturbative Matching
  Between Equal-Time and Lightcone Quantization}},
  \href{http://dx.doi.org/10.1007/JHEP10(2020)092}{\emph{JHEP} {\bf 10} (2020)
  092}, [\href{https://arxiv.org/abs/1812.08177}{{\tt 1812.08177}}].

\bibitem{Anand:2017yij}
N.~Anand, V.~X. Genest, E.~Katz, Z.~U. Khandker and M.~T. Walters, \emph{{RG
  flow from $\phi^4$ theory to the 2D Ising model}},
  \href{http://dx.doi.org/10.1007/JHEP08(2017)056}{\emph{JHEP} {\bf 08} (2017)
  056}, [\href{https://arxiv.org/abs/1704.04500}{{\tt 1704.04500}}].

\bibitem{Chabysheva:2015ynr}
S.~Chabysheva, \emph{{Light-front $\phi^4_{1+1}$ theory using a many-boson
  symmetric-polynomial basis}},
  \href{http://dx.doi.org/10.1007/s00601-016-1106-0}{\emph{Few Body Syst.} {\bf
  57} (2016) 675--680}, [\href{https://arxiv.org/abs/1512.08770}{{\tt
  1512.08770}}].

\bibitem{Burkardt:2016ffk}
M.~Burkardt, S.~S. Chabysheva and J.~R. Hiller, \emph{{Two-dimensional
  light-front $\phi^4$ theory in a symmetric polynomial basis}},
  \href{http://dx.doi.org/10.1103/PhysRevD.94.065006}{\emph{Phys. Rev.} {\bf
  D94} (2016) 065006}, [\href{https://arxiv.org/abs/1607.00026}{{\tt
  1607.00026}}].

\bibitem{Elliott:2014fsa}
B.~Elliott, S.~S. Chabysheva and J.~R. Hiller, \emph{{Application of the
  light-front coupled-cluster method to $\phi^4$ theory in two dimensions}},
  \href{http://dx.doi.org/10.1103/PhysRevD.90.056003}{\emph{Phys. Rev.} {\bf
  D90} (2014) 056003}, [\href{https://arxiv.org/abs/1407.7139}{{\tt
  1407.7139}}].

\bibitem{Chabysheva:2016ehd}
S.~S. Chabysheva and J.~R. Hiller, \emph{{Light-front $\phi_2^4$ theory with
  sector-dependent mass}},
  \href{http://dx.doi.org/10.1103/PhysRevD.95.096016}{\emph{Phys. Rev.} {\bf
  D95} (2017) 096016}, [\href{https://arxiv.org/abs/1612.09331}{{\tt
  1612.09331}}].

\bibitem{Bajnok:2015bgw}
Z.~Bajnok and M.~Lajer, \emph{{Truncated Hilbert space approach to the 2d
  $\phi^{4}$ theory}},
  \href{http://dx.doi.org/10.1007/JHEP10(2016)050}{\emph{JHEP} {\bf 10} (2016)
  050}, [\href{https://arxiv.org/abs/1512.06901}{{\tt 1512.06901}}].

\bibitem{Rychkov:2014eea}
S.~Rychkov and L.~G. Vitale, \emph{{Hamiltonian truncation study of the
  $\phi^4$ theory in two dimensions}},
  \href{http://dx.doi.org/10.1103/PhysRevD.91.085011}{\emph{Phys. Rev.} {\bf
  D91} (2015) 085011}, [\href{https://arxiv.org/abs/1412.3460}{{\tt
  1412.3460}}].

\bibitem{Rychkov:2015vap}
S.~Rychkov and L.~G. Vitale, \emph{{Hamiltonian truncation study of the
  $\phi^4$ theory in two dimensions II. The $\mathbb Z_2$-broken phase and the
  Chang duality}},
  \href{http://dx.doi.org/10.1103/PhysRevD.93.065014}{\emph{Phys. Rev.} {\bf
  D93} (2016) 065014}, [\href{https://arxiv.org/abs/1512.00493}{{\tt
  1512.00493}}].

\bibitem{Elias-Miro:2017xxf}
J.~Elias-Miro, S.~Rychkov and L.~G. Vitale, \emph{{High-Precision Calculations
  in Strongly Coupled Quantum Field Theory with Next-to-Leading-Order
  Renormalized Hamiltonian Truncation}},
  \href{http://dx.doi.org/10.1007/JHEP10(2017)213}{\emph{JHEP} {\bf 10} (2017)
  213}, [\href{https://arxiv.org/abs/1706.06121}{{\tt 1706.06121}}].

\bibitem{Elias-Miro:2017tup}
J.~Elias-Miro, S.~Rychkov and L.~G. Vitale, \emph{{NLO Renormalization in the
  Hamiltonian Truncation}},
  \href{http://dx.doi.org/10.1103/PhysRevD.96.065024}{\emph{Phys. Rev. D} {\bf
  96} (2017) 065024}, [\href{https://arxiv.org/abs/1706.09929}{{\tt
  1706.09929}}].

\bibitem{Elias-Miro:2015bqk}
J.~Elias-Miro, M.~Montull and M.~Riembau, \emph{{The renormalized Hamiltonian
  truncation method in the large $E_T$ expansion}},
  \href{http://dx.doi.org/10.1007/JHEP04(2016)144}{\emph{JHEP} {\bf 04} (2016)
  144}, [\href{https://arxiv.org/abs/1512.05746}{{\tt 1512.05746}}].

\bibitem{truncboot}
{H. Chen, A. L. Fitzpatrick, and D. Karateev}, \emph{{Bootstrapping 2d $\phi^4$
  Theory with Hamiltonian Truncation Data}},
  \href{https://arxiv.org/abs/2107.XXXXX}{{\tt 2107.XXXXX}}.

\bibitem{Pauli:1985ps}
H.~C. Pauli and S.~J. Brodsky, \emph{{Discretized Light Cone Quantization:
  Solution to a Field Theory in One Space One Time Dimensions}},
  \href{http://dx.doi.org/10.1103/PhysRevD.32.2001}{\emph{Phys. Rev.} {\bf D32}
  (1985) 2001}.

\bibitem{Harindranath:1987db}
A.~Harindranath and J.~Vary, \emph{{Solving two-dimensional $\phi^4$ theory by
  discretized light front quantization}},
  \href{http://dx.doi.org/10.1103/PhysRevD.36.1141}{\emph{Phys. Rev. D} {\bf
  36} (1987) 1141--1147}.

\bibitem{Harindranath:1988zt}
A.~Harindranath and J.~Vary, \emph{{Stability of the Vacuum in Scalar Field
  Models in $1 + 1$ Dimensions}},
  \href{http://dx.doi.org/10.1103/PhysRevD.37.1076}{\emph{Phys. Rev. D} {\bf
  37} (1988) 1076--1078}.

\bibitem{Anand:2019lkt}
N.~Anand, Z.~U. Khandker and M.~T. Walters, \emph{{Momentum space CFT
  correlators for Hamiltonian truncation}},
  \href{http://dx.doi.org/10.1007/JHEP10(2020)095}{\emph{JHEP} {\bf 10} (2020)
  095}, [\href{https://arxiv.org/abs/1911.02573}{{\tt 1911.02573}}].

\bibitem{Delacretaz:2018xbn}
L.~V. Delacr\'etaz, A.~L. Fitzpatrick, E.~Katz and L.~G. Vitale,
  \emph{{Conformal Truncation of Chern-Simons Theory at Large $N_f$}},
  \href{http://dx.doi.org/10.1007/JHEP03(2019)107}{\emph{JHEP} {\bf 03} (2019)
  107}, [\href{https://arxiv.org/abs/1811.10612}{{\tt 1811.10612}}].

\bibitem{bender2013advanced}
C.~M. Bender and S.~A. Orszag, \emph{Advanced mathematical methods for
  scientists and engineers I: Asymptotic methods and perturbation theory}.
\newblock Springer Science \& Business Media, 2013.

\bibitem{Hogervorst:2014rta}
M.~Hogervorst, S.~Rychkov and B.~C. van Rees, \emph{{Truncated conformal space
  approach in d dimensions: A cheap alternative to lattice field theory?}},
  \href{http://dx.doi.org/10.1103/PhysRevD.91.025005}{\emph{Phys. Rev. D} {\bf
  91} (2015) 025005}, [\href{https://arxiv.org/abs/1409.1581}{{\tt
  1409.1581}}].

\bibitem{Elias-Miro:2020qwz}
J.~Elias-Mir\'o and E.~Hardy, \emph{{Exploring Hamiltonian Truncation in
  $\bf{d=2+1}$}},
  \href{http://dx.doi.org/10.1103/PhysRevD.102.065001}{\emph{Phys. Rev. D} {\bf
  102} (2020) 065001}, [\href{https://arxiv.org/abs/2003.08405}{{\tt
  2003.08405}}].

\bibitem{Anand:2020qnp}
N.~Anand, E.~Katz, Z.~U. Khandker and M.~T. Walters, \emph{{Nonperturbative
  dynamics of (2+1)d $\phi^4$-theory from Hamiltonian truncation}},
  \href{http://dx.doi.org/10.1007/JHEP05(2021)190}{\emph{JHEP} {\bf 05} (2021)
  190}, [\href{https://arxiv.org/abs/2010.09730}{{\tt 2010.09730}}].

\bibitem{Dempsey:2021xpf}
R.~Dempsey, I.~R. Klebanov and S.~S. Pufu, \emph{{Exact Symmetries and
  Threshold States in Two-Dimensional Models for QCD}},
  \href{https://arxiv.org/abs/2101.05432}{{\tt 2101.05432}}.

\bibitem{Katz:2014uoa}
E.~Katz, G.~Marques~Tavares and Y.~Xu, \emph{{A solution of 2D QCD at Finite
  $N$ using a conformal basis}},  \href{https://arxiv.org/abs/1405.6727}{{\tt
  1405.6727}}.

\bibitem{Katz:2013qua}
E.~Katz, G.~Marques~Tavares and Y.~Xu, \emph{{Solving 2D QCD with an adjoint
  fermion analytically}},
  \href{http://dx.doi.org/10.1007/JHEP05(2014)143}{\emph{JHEP} {\bf 05} (2014)
  143}, [\href{https://arxiv.org/abs/1308.4980}{{\tt 1308.4980}}].

\bibitem{Bhanot:1993xp}
G.~Bhanot, K.~Demeterfi and I.~R. Klebanov, \emph{{(1+1)-dimensional large N
  QCD coupled to adjoint fermions}},
  \href{http://dx.doi.org/10.1103/PhysRevD.48.4980}{\emph{Phys. Rev.} {\bf D48}
  (1993) 4980--4990}, [\href{https://arxiv.org/abs/hep-th/9307111}{{\tt
  hep-th/9307111}}].

\bibitem{Demeterfi:1993rs}
K.~Demeterfi, I.~R. Klebanov and G.~Bhanot, \emph{{Glueball spectrum in a
  ($1+1$)-dimensional model for QCD}},
  \href{http://dx.doi.org/10.1016/0550-3213(94)90236-4}{\emph{Nucl. Phys. B}
  {\bf 418} (1994) 15--29}, [\href{https://arxiv.org/abs/hep-th/9311015}{{\tt
  hep-th/9311015}}].

\bibitem{dalley1993string}
S.~Dalley and I.~R. Klebanov, \emph{{String spectrum of ($1+1$)-dimensional
  large-N QCD with adjoint matter}},
  \href{http://dx.doi.org/10.1103/PhysRevD.47.2517}{\emph{Physical Review D}
  {\bf 47} (1993) 2517}.

\bibitem{Pozsgay:2007kn}
B.~Pozsgay and G.~Takacs, \emph{{Form-factors in finite volume I: Form-factor
  bootstrap and truncated conformal space}},
  \href{http://dx.doi.org/10.1016/j.nuclphysb.2007.06.027}{\emph{Nucl. Phys. B}
  {\bf 788} (2008) 167--208}, [\href{https://arxiv.org/abs/0706.1445}{{\tt
  0706.1445}}].

\bibitem{acerbi1996form}
C.~Acerbi, G.~Mussardo and A.~Valleriani, \emph{Form factors and correlation
  functions of the stress--energy tensor in massive deformation of the minimal
  models (en)}, {\emph{arXiv preprint hep-th/9601113} (1996) }.

\bibitem{Weinberg:1995mt}
S.~Weinberg, \emph{{The Quantum theory of fields. Vol. 1: Foundations}}.
\newblock Cambridge University Press, 2005.

\bibitem{Karateev:2020axc}
D.~Karateev, \emph{{Two-point Functions and Bootstrap Applications in Quantum
  Field Theories}},  \href{https://arxiv.org/abs/2012.08538}{{\tt 2012.08538}}.

\bibitem{Zamolodchikov:1986gt}
A.~B. Zamolodchikov, \emph{{Irreversibility of the Flux of the Renormalization
  Group in a 2D Field Theory}}, {\emph{JETP Lett.} {\bf 43} (1986) 730--732}.

\bibitem{Cardy:1988tj}
J.~L. Cardy, \emph{{The Central Charge and Universal Combinations of Amplitudes
  in Two-dimensional Theories Away From Criticality}},
  \href{http://dx.doi.org/10.1103/PhysRevLett.60.2709}{\emph{Phys. Rev. Lett.}
  {\bf 60} (1988) 2709}.

\bibitem{koekoek2010hypergeometric}
R.~Koekoek, T.~Koornwinder, P.~Lesky and R.~Swarttouw, \emph{Hypergeometric
  Orthogonal Polynomials and Their q-Analogues}.
\newblock Springer Monographs in Mathematics. Springer Berlin Heidelberg, 2010.

\end{thebibliography}\endgroup

\end{document}